\documentclass[journal]{IEEEtran}
\IEEEoverridecommandlockouts
\usepackage{cite}
\usepackage{amsmath,amssymb,amsfonts}
\usepackage{graphicx}
\usepackage{subcaption}
\usepackage{float}
\usepackage{textcomp}
\usepackage{xcolor}
\usepackage[english]{babel}
\usepackage[algo2e]{algorithm2e}
\usepackage{algorithm}
\usepackage{algpseudocode}
\usepackage{filecontents}

\usepackage[font=small,labelfont=bf]{caption}
\def\BibTeX{{\rm B\kern-.05em{\sc i\kern-.025em b}\kern-.08em
		T\kern-.1667em\lower.7ex\hbox{E}\kern-.125emX}}

\def\BibTeX{{\rm B\kern-.05em{\sc i\kern-.025em b}\kern-.08em
    T\kern-.1667em\lower.7ex\hbox{E}\kern-.125emX}}

\newcommand{\bd}{\begin{description}}
\newcommand{\ed}{\end{description}}
\newcommand{\be}{\begin{enumerate}}
\newcommand{\ee}{\end{enumerate}}
\newcommand{\bi}{\begin{itemize}}
\newcommand{\ei}{\end{itemize}}
\newcommand{\bl}{\begin{list}}
\newcommand{\el}{\end{list}}
\newcommand{\bt}{\begin{tabbing}}
\newcommand{\et}{\end{tabbing}}

\definecolor{BLUE}{rgb}{0,0,1}

\usepackage{acronym}  
\acrodef{siso}[SISO]{single-input single-output}
\acrodef{csi}[CSI]{channel state information}
\acrodef{bem}[BEM]{basis expansion model}
\acrodef{ls}[LS]{least squares}
\acrodef{mmse}[MMSE]{minimum mean square error}
\acrodef{pa}[PA]{pilot-aided}
\acrodef{dd}[DD]{decision-directed}
\acrodef{ce}[CE]{channel estimation}
\acrodef{mse}[MSE]{mean square error}
\acrodef{dnn}[DNN]{deep neural network}
\acrodef{mse}[MSE]{mean-squared error}
\acrodef{dl}[DL]{deep learning}
\acrodef{ci}[CI]{channel state information}
\acrodef{mmse}[MMSE]{minimum mean square error}
\acrodef{awgn}[AWGN]{additive white Gaussian noise}
\acrodef{map}[MAP]{maximum a posteriori probability}
\acrodef{ber}[BER]{bit error rate}
\acrodef{kf}[KF]{Kalman filter}
\acrodef{snr}[SNR]{signal-to-noise ratio}
\acrodef{lse}[LSE]{least-squared error}
\acrodef{mmwave}[mmWave]{millimeter wave}
\acrodef{mimo}[MIMO]{multiple-input multiple-output}
\acrodef{uhf}[UHF]{UltraHigh Frequency}
\acrodef{stbc}[STBC]{space-time block code}
\acrodef{nlos}[NLOS]{non-line-of-sight}
\acrodef{st}[ST]{space-time}
\acrodef{ml}[ML]{maximum likelihood}
\acrodef{flops}[Flops]{floating-point operations}
\acrodef{zf}[ZF]{zero forcing}
\acrodef{iir}[IIR]{infinite impulse response}
\acrodef{iir}[IIR]{infinite impulse response}
\acrodef{r}[R]{relaxed}
\acrodef{iid}[IID]{independent and identically distributed}
\acrodef{los}[LOS]{line-of-sight}

\usepackage[yyyymmdd,hhmmss]{datetime} 
\newdateformat{monthyeardate}{\monthname[\THEMONTH] \THEDAY, \THEYEAR} 

\begin{document}
\renewcommand{\figurename}{Fig.}
\title{Decision Directed Channel Estimation Based on Deep Neural Network $k$-step Predictor for MIMO Communications in 5G   \\
	\thanks{M. Mehrabi, M.\ Mohammadkarimi, M. Ardakani, and Y. Jing are with the
		Faculty of Electrical and Computer Engineering, University of Alberta, Edmonton, AB, Canada.
		(e-mail: \texttt{\{mehrtash, mostafa.mohammadkarimi, ardakani, yindi\}@ualberta.ca}).}
}
\author{Mehrtash Mehrabi, \textit{Student Member, IEEE},~Mostafa~Mohammadkarimi, \textit{Member, IEEE},\\~Masoud~Ardakani,~\textit{Senior~Member,~IEEE}, and ~Yindi~Jing,~\textit{Member,~IEEE}}


	\maketitle

	\begin{abstract}
		We consider the use of \ac{dnn} to develop a \ac{dd}-\ac{ce} algorithm
       for \ac{mimo}-space-time block coded systems in highly dynamic vehicular environments. We propose the use of \ac{dnn} for $k$-step channel prediction for \ac{stbc}s, and show that \ac{dl}-based \ac{dd}-\ac{ce} can removes the need for Doppler spread estimation in fast time-varying quasi stationary channels, where the Doppler spread varies from one packet to another. Doppler spread estimation in this kind of vehicular channels is remarkably challenging and requires a large number of pilots and preambles, leading to lower power and spectral efficiency.
       We train two \ac{dnn}s which learn 
       \
       real and imaginary parts of the \ac{mimo} fading channels over a wide range of Doppler spreads.
       We demonstrate that by those \ac{dnn}s, \ac{dd}-\ac{ce} can be realized with only rough priori knowledge about Doppler spread range.
       For the proposed DD-CE algorithm, we also analytically derive the maximum likelihood (ML) decoding algorithm for STBC transmission.
       The proposed \ac{dl}-based \ac{dd}-\ac{ce} is a promising solution for reliable communication over the vehicular \ac{mimo} fading channels without accurate mathematical models.
       This is because \ac{dnn} can intelligently learn the statistics of the fading channels.
       Our simulation results show that the proposed \ac{dl}-based \ac{dd}-\ac{ce} algorithm exhibits  lower propagation error compared to existing \ac{dd}-\ac{ce} algorithms while the latters require perfect knowledge of the Doppler rate.
	\end{abstract}

	\begin{IEEEkeywords}
		MIMO communication, channel estimation, deep learning, decision directed, mmWave communications.
	\end{IEEEkeywords}
	
	%
	\IEEEpeerreviewmaketitle
	
	\section{Introduction}
	\subsection{Motivation}
	\acresetall
	\IEEEPARstart{W}{ireless}
	data traffic has been growing rapidly and according to a Cisco report \cite{cisco}, global wireless data traffic will increase sevenfold between 2016 and 2021, reaching 49.0 exabytes per month by 2021.
	This wireless data explosion is expected to accelerate over the next decade by increasing  the popularity of smartphones, continual use of wireless video streaming services, and the rise of the Internet-of-Things (IoT) \cite{IoT,mmWave_5G}.
	
	In order to address this high volume of data traffic demand, one promising solution is to enable the use of higher frequency spectrum, e.g., \ac{mmwave} \cite{mmWave_5G,mmWave_5G_2}.
	Currently, the fifth-generation (5G) wireless communications is being developed based on utilizing \ac{mmwave} frequencies (30-300 GHz) to provide a notable spectrum and high data rates on the order of Gbps \cite{mmWave_5G_2}.
	Utilizing \ac{mmwave} in \ac{mimo} communication systems is one of the candidate technologies for 5G wireless standardization \cite{mimo_mmwave} which can further improve the system capacity.
	It also benefits from spatial diversity against small-scale fading, higher data rates, and the ability to cancel interference \cite{mimo1,mimo2,mimo3}. In addition to spatial diversity due to multiple transmit and receive antennas, time diversity can also be achieved through \ac{stbc}.
\ac{stbc} is an advanced transmission technique used in multiple antenna systems to transmit multiple replicas of information symbols to exploit the various received versions of the transmitted symbols to improve the reliability of transmission.

	In the design of a \ac{mimo} wireless communication system in 5G, there are two main challenges that affect the performance, namely channel modeling and \ac{ce}.
	Due to the complex propagation characteristics of highly dynamic channels, channel modeling is an extremely challenging task \cite{ChannelModeling}.
	Furthermore, in highly dynamic environments, the channel impulse response varies quickly, and thus, the channel statistics remain constant only for a very short period of time.
	Consequently, the high channel variations limits the channel modeling and reduces the performance of existing channel estimators \cite{highSpeedComm1}.
	These constraints on \ac{ce} in highly dynamic environments are even more crucial for \ac{stbc} transmission where more channels must be estimated for each block transmission and its associated decoding process is considerably affected by the accuracy of \ac{ce} method.

	For \ac{mimo} systems, \ac{ce} schemes have been mostly based on pilot-assisted approaches, assuming a quasi-static fading model that allows the channel to be constant for a block of symbols and change independently to a new realization.
	This is not applicable for environments such as fast time-varying channels where the coherence time is considerably short.
	Currently, \ac{dd}-\ac{ce} methods have been suggested to be employed in time-varying channels and it has been widely used in vehicular communication systems based on IEEE 802.11p technology \cite{dd_pilot}.
	In \ac{dd}-\ac{ce}, first a block of training symbols is sent to estimate the \ac{csi}.
	Then, data transmissions are conducted, where the subsequent \ac{csi} corresponding to the data symbols are predicted by treating the detected symbols as training data and re-estimating the channel iteratively \cite{recursiveLS,Decisiondirected}.
	The core part of the \ac{dd}-\ac{ce} is channel prediction.
	Existing channel predictors are highly depended on channel statistics which is severely affected by the estimation of Doppler spread of the channel.
	However, in highly dynamic vehicular environments, Doppler spread estimation is challenging.

	Recently, \ac{dl} has been widely investigated in the signal processing and communications problems to improve the performance of some certain parts of conventional communication systems, such as decoding, estimation, and more  \cite{ye2018power,nachmani2018deep,farsad2017detection,kim2018deep,o2017introduction,dorner2018deep,o2017deep,DLSphere}.
	In particular, \ac{dl}-based \ac{ce} methods have been studied in literature such as the recent work in \cite{ye2018power}.
	A \ac{dnn} is a universal function approximator with superior logarithmic learning ability and convenient optimization capability, and thus can be used for the problems without any accurate mathematical model \cite{deeplearning}.
	Currently, most of the existing algorithms in communications rely on precise mathematical models.
	However, in practice tractable mathematical models cannot reflect many imperfections and nonlinearities, and can only work as rough approximations when these issues are non-negligible.
	\ac{dl} can fix this drawback in communication and information theory and offer algorithms without mathematically tractable models \cite{o2017introduction}.

	Motivated by the limitations of existing channel predictors and the strength of \ac{dnn} in learning and prediction, a \ac{dl}-based \ac{dd}-\ac{ce} for \ac{mimo} \ac{stbc} is proposed in this paper, where the \ac{mimo} channel coefficients are predicted by two trained \ac{dnn}s.
	While existing channel predictors require the exact value of Doppler spread and an accurate mathematical model for Doppler spectrum, our proposed algorithm does not require  Doppler spread estimation and provides a more reliable packet transmission in highly dynamic vehicular environments.
	Moreover, we derive the \ac{ml} \ac{stbc} decoding for any \ac{stbc} design in fast time-varying channels, where channels vary during each \ac{stbc} transmission.
	In the proposed scheme, first we predict the corresponding channels for each block transmission and then perform signal detection with the channel prediction.

	\subsection{Related Work}
	\textit{1) DD-CE channel predictors:}
	The optimal Weiner filter, finite length Wiener filter, and  weighted recursive \ac{ls}, such as \ac{kf}, are the most popular predictors employed in \ac{dd}-\ac{ce} \cite{Kalmanfilter,Kalman2} where all of them rely on exact Doppler spread estimation.
	Furthermore, inaccurate Doppler spread estimation results in significant error propagation in suboptimal channel predictors like \ac{kf}, especially at high Doppler spreads and when the transmitted packets are large.
	In addition, the sensitivity of current channel predictors to the accuracy of the channel model is too high to tolerate any modeling errors.
	However, finding an explicit mathematical model to describe the channel propagation characteristics in highly dynamic environments is a challenging task and thus modeling error is inevitable.
	
	\textit{2) \ac{ce} for \ac{stbc} transmission:}
 	Wireless communication systems usually rely on some form of diversity at the transmit side and/or the receiver side.
 	\ac{stbc} transmission is one of the most common solutions for achieving diversity.
 	Alamouti introduced a well-known transmission technique for systems with two transmit antennas in \cite{alamouti}.
 	By generalizing Alamouti's idea, Tarokh \textit{et al.} proposed \ac{stbc} for other numbers of transmit/receive antennas \cite{tarokh2}.
 	The problem of \ac{ce} when \ac{stbc} is used for transmission has been investigated in many studies \cite{ceKalman2,Gia_method,tarokh_cestbc}.
 	In the current studies, whenever \ac{stbc} is used in a time-varying channel, two approaches were employed.
	One of them is considering a coherent channel for block transmission and the other one is channel modeling with a rough approximation such as first order autoregressive model.	
	In \cite{alamouti,tarokh_cestbc}, the authors assumed that the channel is coherent for each block transmission and using this assumption, they proposed a coherent detection algorithm.
	However, in the fast time-varying channel where the channel statistics change rapidly, all the mentioned assumptions lead to a degraded signal detection.
 	A \ac{kf}-based \ac{ce} method was used in \cite{ceKalman2,Gia_method,ceKalman1} and it was assumed that for a block transmission the fading channel is changed based on first order Gauss-Markov process. Then after determining the channels for each block transmission, they developed a detection algorithm.
 	Furthermore, the proposed detection algorithms are only valid for Alamouti's scheme and for longer \ac{stbc} block transmission the assumptions are not applicable \cite{Gia_method,tarokh_cestbc,ceKalman2}.


	\subsection{Contribution}

The main contributions of this papers are as follows:

	\begin{itemize}
\item We propose a \ac{dl}-based $k$-step channel predictor;
\item A new \ac{dd}-\ac{ce} algorithm based on the proposed predictor is proposed  for \ac{mimo}-\ac{stbc} systems. The proposed algorithm exhibits the following advantages:
\begin{itemize}
\item It removes the need for Doppler spread estimation;
\item It exhibits lower error propagation compared to existing algorithm;
\item It can be applied to \ac{mimo} fading channels without concrete mathematical models;
\item It has a lower computational complexity compared to existing \ac{dd}-\ac{ce} algorithms;
\item It is applicable to even large packets;
\end{itemize}
\item The joint \ac{ml} decoding algorithm for general \ac{stbc}s in time-varying fading channels is derived;
\item The proposed scheme has better performance than existing algorithms.
\item We derive the optimal \ac{dd}-\ac{ce} for general \ac{stbc}s using Wiener predictor.
\end{itemize}

	\subsection{Organization}
	The outline of this paper is as follows.
	In Section \ref{sec:DL}, we briefly review \ac{dl}.
	Section~\ref{sec:system} presents the system model. Section~\ref{sec:ce} introduces \ac{dd}-\ac{ce} method for \ac{mimo} wireless communications.
	Section~\ref{sec:DL_CE} describes the proposed \ac{dl}-based \ac{dd}-\ac{ce} algorithm along with our proposed \ac{ml} decoding algorithm for \ac{stbc} design .
	The complexity analysis of our proposed algorithm is presented in Section \ref{sec:comp}.
	Simulation results are provided in Section~\ref{sec:sim}, and finally we conclude the paper in Section~\ref{sec:con}.
	
	\subsection{Notation}
	Throughout this paper, $(\cdot)^*$ represents the complex conjugate. The real and imaginary parts of a complex number are denoted by $\Re\{\cdot\}$ and $\Im\{\cdot\}$, respectively.
	Matrix transpose and Hermitian operators are shown by $(\cdot)^T$ and $(\cdot)^H$, respectively.
	Moreover, the inverse of matrix $\mathbf{A}$ is represented by $\mathbf{A}^{-1}$ and the symbol $\mathbf{I}_w$ denotes the identity matrix of size $w$.
	The column vector of size $z$ and all ones is denoted by $\mathbf{1}_z$.
	The operator $\text{diag}(\mathbf{b})$ returns a square diagonal matrix with the elements of vector $\mathbf{b}$ on the main diagonal.
	Assuming $\mathbf{a}, \mathbf{b}, \mathbf{c}$ and $\mathbf{d}$ are some matrices with whether equal or different sizes, operator $\text{bdiag}(\mathbf{a},\mathbf{b},\mathbf{c},\mathbf{d})$ returns the following matrix
	\begin{equation}
		\begin{bmatrix}
			\mathbf{a} & 0 & 0 & 0 \\
			0 & \mathbf{b} & 0 & 0 \\
			0 & 0 & \mathbf{c} & 0 \\
			0& 0 & 0 & \mathbf{d}
		\end{bmatrix}.
	\end{equation}
	Furthermore, $|\cdot|$ shows the absolute value, $\mathbb{E}\{\cdot\}$ is the statistical expectation, $\hat{\mathbf{a}}$ denotes an estimated value for vector $\mathbf{a}$, and the Frobenius norm of vector $\mathbf{a}$ is showed by $\|\mathbf{a}\|$.
	The constellation and $m$-dimensional complex spaces are denoted by $\mathbb{D}$ and $\mathbb{C}^m$, respectively.
	For the sake of simplicity, the element-wise notation of Matlab is used, where $\mathbf{A}_{:,k_1:k_2}$ denotes all rows and columns $k_1,k_1+1,\cdots,k_2$ of matrix $\mathbf{A}$, and
the notation $\mathbf{a}_{k_1:k_2}$ shows the $k_1$-th until $k_2$-th entries of vector $\mathbf{a}$. Note that in the sequel, $\mathbf{A}_k$ and $\mathbf{a}_k$ represents $k$-th matrix and $k$-th vector, respectively.
	Finally, the circularly symmetric complex Gaussian distribution with mean vector $\mathbf{\mu}$ and covariance matrix $\mathbf{\Sigma}$ is denoted by $\mathcal{CN}(\mathbf{\mu},\mathbf{\Sigma})$.
	
	\section{Deep Learning}\label{sec:DL}
	\begin{figure}[]
	\centering
	\includegraphics[width=.45\textwidth]{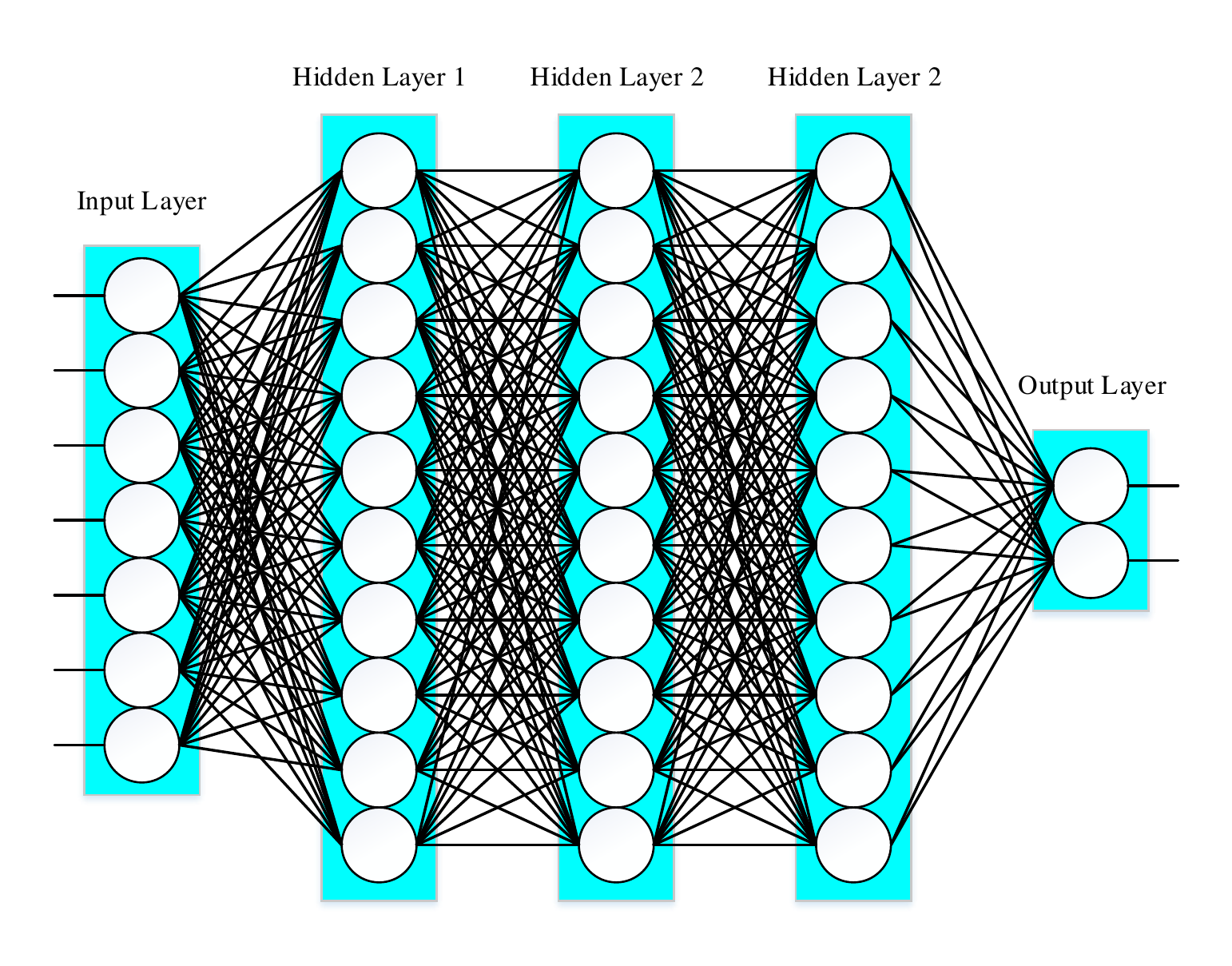}
	\caption{Typical \ac{dnn} with three hidden layers.}
	\label{fig_dnn}
	\vspace{-1em}
    \end{figure}
	
	Deep learning is an approach to artificial intelligence and more specifically, it is a type of machine learning technique that enables computer systems to learn complicated concepts without any need for exact mathematical operators.
	As a result, computer systems can learn from a series of experiences to find a solution by a hierarchy of concepts where each concept is defined by simpler concepts.
	These concepts on top of each other, generate a deep graph to show the mapping between input and output and because of the depth of this graph this approach to artificial intelligence is called deep learning \cite{deeplearning}.
	
	One of the quintessential \ac{dl} models are deep feedforward networks, also called deep neural networks, where by training a vector of learning parameters $\mathbf{\Theta}$, some function $f$ is approximated as
	\begin{equation} \label{eq:dnnfunc}
		\mathbf{y} = f(\mathbf{x};\mathbf{\Theta}),
	\end{equation}
	where the input vector $\mathbf{x}\in\mathbb R^n$ is mapped to the output vector $\mathbf{y}\in\mathbb R^m$.
	The \ac{dnn} breaks this complicated mapping into a series of simple ones, each defined by a distinct layer of \ac{dnn}.
	A \ac{dnn} is built by a sequence of visible and hidden layers.
	At visible layers, we are able to observe the variables.
	The input and output layers of a \ac{dnn} are both visible layers.
	At hidden layers, the variables are not accessible and their values are changed based on the feature extraction.
	Fig. \ref{fig_dnn} shows a \ac{dnn} with three hidden layers.
	For example, in a \ac{dnn} with $L$ hidden layers, we can represent function $f$ in \eqref{eq:dnnfunc} by $L$ functions $f^{(1)}, f^{(2)}, \cdots, f^{(L)}$ as
	\begin{equation}
		\mathbf{y} \approx f^{(L)}\Big(\cdots f^{(2)}\big(f^{(1)}(\mathbf{x};\Theta_1);\Theta_2\big)\cdots;\Theta_L\Big).
	\end{equation}
	Each function $f^{(l)}$ is defined as
	\begin{equation}
		f^{(l)}(\mathbf{z};\Theta_l) \triangleq A_l(\mathbf{w}_l\mathbf{z}+\mathbf{b}_l), ~~~~~ l=1, 2, \cdots, L
	\end{equation}
	where $\mathbf{z}$ is the output of the previous layer, $\mathbf{\Theta}_l\triangleq\{\mathbf{w}_l,\mathbf{b}_l\}$ denotes the set of learning parameters, $\mathbf{w}_l\in\mathbb{R}^{n_l\times n_{l-1}}$ and $\mathbf{b}_l\in\mathbb{R}^{n_l}$ ($n_0=n$ and $n_L=m$) represent weights and biases, respectively and $A_l$ is the activation function of the $l$-th layer.
	By training the \ac{dnn} with a training set and a known desired output, the weights and biases can be learned \cite{deeplearning}.

	\section{System Model}\label{sec:system}
	We consider a \ac{mimo} system in a time-varying flat fading channel, where the transmitter and receiver are equipped with $n_{\rm{t}}$ and $n_{\rm{r}}$ antennas.
	The space-time encoder at the transmitter takes a block ${\bf{s}}_i \in \mathbb{D}^{{N}_{{\rm{s}}}}$ of ${{N}_{{\rm{s}}}}$ information symbols as input and maps it into a \ac{stbc} matrix $\bar{\bf{C}}_i$ as
	\begin{equation}
	\bar{\bf{C}}_i \triangleq
	\begin{bmatrix}
	c_{11} & c_{12} & \cdots & c_{1n_{\rm{x}}} \\
	c_{21} & c_{22} & \cdots & c_{2n_{\rm{x}}} \\
	\vdots & \vdots & & \vdots \\
	c_{n_{\rm{t}}1} & c_{n_{\rm{t}}2} & \cdots & c_{n_{\rm{t}}n_{\rm{x}}}
	\end{bmatrix},
	\end{equation}
	where $\mathbb{D}$ is an arbitrary constellation and $c_{pq}$, $p=1,\cdots,n_t$ and $q=1,\cdots,n_x$ are functions of the information vector ${\bf{s}}_i$.
	The $n_{\rm{x}}$ columns of $\bar{\bf{C}}_i$ are generated in $n_{\rm{x}}$ successive time intervals each of duration $T_{\rm{s}}$, while each of the $n_{\rm{t}}$ entries in a given column is forwarded to one of the $n_{\rm{t}}$ transmit antennas. At the $m$-th transmit antenna, $c_{mk}$ is first  pulse shaped and then transmitted during the $k$-th time interval. The transmitted waveforms from $n_{\rm{t}}$ transmit antennas are sent simultaneously.

		\begin{figure*}
		\setcounter{equation}{13}
		\begin{align} \label{03vzn3099vb41}
		\hat{\mathbf{h}}_{k|k-1}^{(n)}&={\mathbf{\Sigma}}_{\mathbf{h}_{k}^{(n)},{\bf{y}}_{1:k-1}^{(n)}}\mathbf{\Sigma}_{{\bf{y}}_{1:k-1}^{(n)}}^{-1}{\bf{y}}_{1:k-1}^{(n)}
		={{\mathbf{A}}}_{k-1}^{\rm{H}}{\mathbf{U}}_{k-1}^{\rm{H}}
		\Big{(}{\mathbf{U}}_{k-1}{{\mathbf{R}}}_{k-1}^{\rm{d}}
		{\mathbf{U}}_{k-1}^{\rm{H}}+\sigma_{\rm{w}}^2\mathbf{I}_{k-1}\Big{)}^{-1}{\mathbf{y}}_{1:k-1}^{(n)},
		\end{align}
		\hrulefill
	\end{figure*}
	\setcounter{equation}{5}
	
	If $n_{\rm{x}}=1$ and $N_{\rm{s}}=n_{\rm{t}}$, independent information symbols are transmitted over each transmit antenna at each time interval, one word.
	This transmission scheme is referred to as the simplest case of spatial multiplexing without any need for precoding and maps a block ${\bf{s}}_i \in \mathbb{D}^{{N}_{{\rm{s}}}}$ of ${{N}_{{\rm{s}}}}$ information symbols to the transmit antennas as
	\begin{equation}
	\bar{\bf{C}}_i \triangleq
	\begin{bmatrix}
	c_{1}  \\
	c_{2}    \\
	\vdots  \\
	c_{n_{\rm{t}}}
	\end{bmatrix}={\bf{s}}_i.
	\end{equation}

	Let us represent the time-varying fading channels between the $n$-th receive antenna and all $n_{\rm{t}}$ transmit antennas at the $k$-th time index (index $k$ is assigned to a continues-time index $t_k=kT_s$) by
	\begin{equation}
	\mathbf{h}_{k}^{(n)}=\begin{bmatrix}
	h_{n1,k} & h_{n2,k} & \dots  & h_{nn_{\rm{t}},k}
	\end{bmatrix}^T,
	\end{equation}
	where $h_{nm,k}$ is the fading channel between the $m$-th transmit antenna and the $n$-th receive antenna at the $k$-th time index.
	It is assumed that the fading channels are independent for different transmit-receive antenna pairs and can be modeled as a wide sense stationary process over the packet time with unknown Doppler rate $\rho \in [\rho_{\rm{min}} , \rho_{\rm{max}}]$ due to the highly dynamic vehicular environments. The autocorrelation function of the complex fading channel between the $m$-th transmit and $n$-th receive antenna over the packet time is modeled as
	\begin{align}\label{eq:DopplerSpread}
	\mathbb{E}\{h_{nm,k_1}h_{nm,k_2}^*\}=R\big(\rho (k_1-k_2)\big)~~~~ \rho \in [\rho_{\rm{min}} , \rho_{\rm{max}}],
	\end{align}
	where $R(\cdot)$ denotes the Doppler spectrum model.
	Widely used ones include the Jakes, Asymmetric Jakes, Gaussain, and flat model \cite{hlawatsch2011wireless}.
	It should be noted that our proposed algorithm does not require any priori knowledge about the Doppler spectrum  model and it is effective even without any explicit mathematical representation for the Doppler spectrum.
	
	We assume that $n_{\rm{b}}$ blocks of \ac{stbc}s are transmitted over a packet of length $(n_{\rm{b}}n_{\rm{x}}+n_{\rm{p}})T_{\rm{s}}$ after the transmission of the pilot matrix ${\bf{P}}$ as
	\begin{align}\label{packet}
	{\bf{C}}\triangleq \begin{bmatrix}
	{\bf{P}} & \bar{\bf{C}}_1 & \bar{\bf{C}}_2 & \dots  & \bar{\bf{C}}_{n_{\rm{b}}}
	\end{bmatrix},
	\end{align}
	where ${\bf{P}}$ is a $n_{\rm{t}}\times n_{\rm{p}}$ orthogonal matrix.
	
	At the receiver, the vector of received baseband signal for the pilot matrix and $n_{\rm{b}}$ transmitted \ac{stbc}s in the packet at the $n$-th received antenna is expressed as
	\begin{equation}\label{eq:system_model}
	\mathbf{y}^{(n)} \hspace{-0.2em} \triangleq  \hspace{-0.2em}
	\begin{bmatrix}
	{{y}_{1}^{(n)}}\\
	{{y}_{2}^{(n)}}\\
	\vdots \\
	{{y}_{L}^{(n)}}
	\end{bmatrix} \hspace{-0.2em}  =\hspace{-0.2em}
	\begin{bmatrix}
	{\mathbf{C}}_{:,1}^T & 0 & 0 & \dots   \\
	0 & \mathbf{C}_{:,2}^T & 0 & \dots   \\
	\vdots & \vdots & \ddots & \vdots  \\
	0 & 0  & \dots  & {\mathbf{C}}_{:,L}^T
	\end{bmatrix} 	
	\begin{bmatrix}
	{\mathbf{h}_{1}^{(n)}}\\
	{\mathbf{h}_{2}^{(n)}} \\
	\vdots \\
	{\mathbf{h}_{L}^{(n)}}
	\end{bmatrix}
	+
	\begin{bmatrix}
	{{w}_{1}^{(n)}}\\
	{{w}_{2}^{(n)}}\\
	\vdots \\
	{{w}_{L}^{(n)}}
	\end{bmatrix}
	\end{equation}
	where $n=1,2,\cdots,n_{\rm{r}}$ and $L \triangleq
	n_{\rm{b}}n_{\rm{x}}+n_{\rm{p}}$. The additive noise vector at the $n$-th receive antennas, i.e., $\mathbf{w}^{(n)}\triangleq [w_1^{(n)},~w_2^{(n)},~\cdots,w_L^{(n)}]^T$ can be either Gaussian or non-Gaussian.
	
	
	

	\section{Decision Directed Channel Estimation for MIMO Communications}\label{sec:ce}
	The core part of the \ac{dd}-\ac{ce} is channel prediction, where the current channel state is estimated based on the previous estimates and detected symbols.
	Under jointly Gaussian dynamic parameters, i.e, noise and fading channel, the optimal channel predictor is the Wiener-type predictor.
	In this section, we derive the optimal one-step and $n_{\rm{x}}$-steps channel prediction for spatial multiplexing and \ac{stbc} transmission, respectively.
	We show that the \ac{dd}-\ac{ce} developed based on the optimal Wiener-type predictor and Kalman filter requires a priori knowledge about the exact Doppler spread, which is extremely difficult to track in highly dynamic environments. Moreover, these estimators surfer from huge computational complexity.

	\subsection{\ac{dd}-\ac{ce} for Spatial Multiplexing Using Wiener Predictor}
	In this subsection, we obtain the \ac{dd}-\ac{ce} for spatial multiplexing transmission by using one-step optimal Wiener predictor and Kalman filter.
	
	\subsubsection{\ac{dd}-\ac{ce} Based on Optimal Wiener Predictor}
	\ac{dd}-\ac{ce} for spatial multiplexing is developed on the basis of one-step channel prediction.
	By employing the optimal one-step Wiener predictor, \ac{dd}-\ac{ce} for spatial multiplexing is expressed as
	\begin{align}\label{77766ggggb777uu90bRR2}
	\hat{\mathbf{h}}_{k|k-1}^{(n)}=\mathbb{E}\big{\{}\mathbf{h}_{k}^{(n)}|{{\mathbf{y}}}_{1:k-1}^{(n)},{\hat{\mathbf{C}}_{:,1:k-1}}\big{\}},\,\ {n_{\rm{p}}+1 \leq k\leq L}
	\end{align}
	where $\mathbf{y}^{(n)}$,
	$n=1,2,\cdots,n_{\rm{r}}$, is given in \eqref{eq:system_model}, and
	\begin{equation}\label{uuubnkop}
	\hat{\mathbf{C}}_{:,k}=
	\begin{cases}
	T_{\rm{sp}}\big{(}{{\mathbf{y}}}_{k},\hat{\mathbf{H}}_{k}\big{)} &  n_{\rm{p}}+1 \leq k\leq L \\
	\mathbf{P}_{:,k}& 1 \leq k\leq n_{\rm{p}}
	\end{cases}.
	\end{equation}
	In \eqref{uuubnkop}, $T_{\rm{sp}}$ can be either the optimal \ac{ml} detector or a suboptimal detector, such as \ac{zf} or \ac{mmse} and uses all the channel estimations and received signals at the $k$-th time index, which are
\begin{subequations}
	\begin{align}
	\hat{\mathbf{H}}_{k} &\triangleq \Big{[}\hat{\mathbf{h}}_{k|k-1}^{(1)}~ \hat{\mathbf{h}}_{k|k-1}^{(2)} ~ \cdots \hat{\mathbf{h}}_{k|k-1}^{(n_{\rm{r}})}\Big{]}^T \\
	{\mathbf{y}}_{k} &\triangleq \Big{[}y_k^{(1)}~ y_k^{(2)} ~ \cdots ~ y_k^{(n_{\rm{r}})}\Big{]}^T.
	\end{align}
\end{subequations}

	\begin{figure*}
	\setcounter{equation}{16}
	\begin{align}\label{7777777777774444444222222222221}
	{\mathbf{\Sigma}}_{k|k-1}^{(n)} & \hspace{-0.3em}\triangleq \hspace{-0.2em} \mathbb{E}\Big{\{}(\hat{\mathbf{h}}_{k|k-1}^{(n)}\hspace{-0.1em}-\hspace{-0.1em}{\mathbf{h}}_{k}^{(n)})(\hat{\mathbf{h}}_{k|k-1}^{(n)}\hspace{-0.1em}-\hspace{-0.1em}{\mathbf{h}}_{k}^{(n)})^{\rm{H}}\big{|}{{\mathbf{y}}}_{1:k-1}^{(n)}\Big{\}}
	={\mathbf{I}}_{n_{\rm{t}}}\hspace{-0.1em}-\hspace{-0.1em}{{\mathbf{A}}}_{k-1}^{\rm{H}}{\mathbf{U}}_{k-1}^{\rm{H}}\Big{(}{\mathbf{U}}_{k-1}{{\mathbf{R}}}_{1:k-1}^{\rm{d}}
	{\mathbf{U}}_{k-1}^{\rm{H}}+\sigma_{\rm{w}}^2\mathbf{I}_{k-1}\Big{)}^{-1}
	{\mathbf{U}}_{k-1}{{\mathbf{A}}}_{k-1}
	\end{align}
	\setcounter{equation}{20}	
	\begin{align}\label{555mm3n3ll11}
	&{\mathbf{\Sigma}}_{k|k-1}^{(n)}=R^2(\rho) \Big{(}{\mathbf{\Sigma}}_{k-1|k-2}^{(n)}-{\mathbf{\Sigma}}_{k-1|k-2}^{(n)}\hat{\mathbf{C}}_{:,1:k-1}
	\Big{(}\hat{\mathbf{C}}_{:,1:k-1}^{\rm{H}}{\mathbf{\Sigma}}_{k-1|k-2}^{(n)}\hat{\mathbf{C}}_{:,1:k-1}+\sigma_{\rm{w}}^2\mathbf{I}_{k-1}\Big{)}^{-1}\hat{\mathbf{C}}_{:,1:k-1}^{\rm{H}}
	{\mathbf{\Sigma}}_{k-1|k-2}^{(n)}\Big{)}
	\end{align}
	\hrulefill
	\end{figure*}
	\setcounter{equation}{14}
			
	For fading channels with zero-mean circular complex Gaussian distribution (i.e., Rayleigh fading channel) and \ac{awgn} at the receiver, the optimal one-step channel predictor in \eqref{77766ggggb777uu90bRR2} for the $n$-th receive antenna is given in \eqref{03vzn3099vb41},
	where\\
	${\mathbf{\Sigma}}_{\mathbf{h}_{k}^{(n)},{\bf{y}}_{1:k-1}^{(n)}}  \hspace{-0.2em}\triangleq \hspace{-0.2em} \mathbb{E} \big{\{}{\mathbf{h}}_{k}^{(n)}{{\bf{y}}_{1:k-1}^{(n){\rm{H}}}}\big{\}}$,
	${\mathbf{\Sigma}}_{{\bf{y}}_{1:k}^{(n)}}  \hspace{-0.2em}\triangleq\hspace{-0.2em} \mathbb{E} \big{\{} {{\bf{y}}_{1:k}^{(n)}}
	{{\bf{y}}_{1:k}^{(n){\rm{H}}}}\big{\}}$,
	
	\hspace{-1em}${\mathbf{U}}_{k-1} \hspace{-0.2em}\triangleq \hspace{-0.2em}$
	$\Big{[}{\rm{diag}}\big{(}{\hat{\mathbf{C}}}_{1,1:k-1}\big{)}
	\hspace{0.1em} {\rm{diag}}\big{(}{\hat{\mathbf{C}}}_{2,1:k-1}\big{)} \cdots {\rm{diag}}\big{(}{\hat{\mathbf{C}}}_{n_{\rm{t}},1:k-1}\big{)}
	\Big{]}$,
	${\mathbf{A}}_{k-1}\triangleq {\mathbf{I}}_{n_{\rm{t}}} \otimes{\mathbf{r}}{(1,k-1)}$, ${\mathbf{R}}_{k-1}^{\rm{d}}={\mathbf{I}}_{n_{\rm{t}}}\otimes {\mathbf{R}}_{k-1}$,
	\begin{align}\label{vvooo90}
	{\mathbf{r}}{(u,v)} \triangleq \Big{[}{R}{(\rho v)} ~ {R}{(\rho(v-1))} ~ \cdots ~  {R}({\rho u})\Big{]}^T,
	\end{align}
	and
	\begin{align}\label{corrttt}
	\hspace{-0.2em}{\mathbf{R}}_{k-1}\hspace{-0.2em}\triangleq \hspace{-0.2em}
	\begin{bmatrix}
	{R}{(0)} & {R}{(\rho)} & \cdots & {R}{(\rho(k-2))}\\
	{R}{(\rho)}  & {R}{(0)} & \cdots & {R}{(\rho(k-3))}\\
	\vdots & \vdots & \cdots & \vdots\\
	{R}{(\rho(k-2))} & {R}{(\rho(k-3))} & \cdots & {R}{(0)}
	\end{bmatrix}.
	\end{align}
	The \ac{mmse} of the optimal one-step channel predictor for spatial multiplexing transmission at the $k$-th time index is given as \eqref{7777777777774444444222222222221}.
	
	As seen, the Weiner filter predictor in \eqref{03vzn3099vb41}
	requires a priori knowledge about the channel statistics through matrixes ${{\mathbf{A}}}_{k-1}$ and
	${{\mathbf{R}}}_{k-1}^{\rm{d}}$.
	However, these statistics vary
	with the Doppler spread of the fading channel $\rho$.
	Hence, Doppler spread estimation prior to CE is required.
	Moreover, the optimal channel predictor suffers from high
	computational complexity due to the matrix inversion in \eqref{03vzn3099vb41}.
	The matrix inversion for the latter symbols of the packet becomes more complex
	due to the higher matrix size.
	Hence, in practice,
	a Weiner filter of order $n_{\rm{p}}$ is employed for
	one-step channel prediction in spatial multiplexing transmission to reduce the complexity.
	
	For the reduced complexity one-step prediction using the Weiner filter of order $n_{\rm{p}}$,
	${{\mathbf{y}}}_{1:k-1}^{(n)}$ and ${\hat{\mathbf{C}}_{n,1:k-1}}$, $n=1,2, \cdots, $ $ n_{\rm{t}}$,
	in \eqref{77766ggggb777uu90bRR2} and \eqref{03vzn3099vb41} are respectively replaced with ${{\mathbf{y}}}_{k-n_{\rm{p}}:k-1}^{(n)}$ and ${\hat{\mathbf{C}}_{n,k-n_{\rm{p}}:k-1}}$.
	The correlation matrix ${{\mathbf{R}}}_{k-1}$ and ${\mathbf{r}}{(1,k-1)}$  are replaced with ${{\mathbf{R}}}_{n_{\rm{p}}}$ and ${\mathbf{r}}{(k-n_{\rm{p}},k-1)}$.
	Moreover, ${\mathbf{U}}_{k-1}$ is modified as ${\mathbf{U}}_{k-1}=\Big{[}{\rm{diag}}\big{(}\hat{\mathbf{C}}_{1,k-n_{\rm{p}}:k-1}\big{)} ~ {\rm{diag}}\big{(}\hat{\mathbf{C}}_{2,k-n_{\rm{p}}:k-1}\big{)} ~ \cdots ~ {\rm{diag}}\big{(}\hat{\mathbf{C}}_{n_{\rm{t}},k-n_{\rm{p}}:k-1}\big{)}\Big{]}$.

	\subsubsection{\ac{dd}-\ac{ce} Based on Kalman Filter}
	For the \ac{mimo} fading channels, where the dynamics of the fading
	process can be molded by a state-space
	Gauss-Markov process as
	\setcounter{equation}{17}
	\begin{align}
	\mathbf{h}_{k+1}^{(n)}=R(\rho)\mathbf{h}_{k}^{(n)}+\mathbf{v}_{k}^{(n)},\,\,\,\,n=1,2,\cdots,n_{\rm{r}},
	\end{align}
	the optimal one-step predictor is a Kalman filter. In this case, \ac{dd}-\ac{ce} for spatial multiplexing transmission can be achieved through an \ac{iir} filter as
	\begin{align}\label{77766ggggb777uu90bee}
	\hat{\mathbf{h}}_{k|k-1}^{(n)}&\hspace{-0.1em}= \hspace{-0.1em}\mathbb{E}\big{\{}\mathbf{h}_{k}^{(n)}|{{\mathbf{y}}}_{1:k-1}^{(n)},{\hat{\mathbf{C}}_{:,1:k-1}}\big{\}} \,\,\,\,\,\ n_{\rm{p}}+1 \leq k \leq L \\ \nonumber
	&=\big{(}R(\rho)\mathbf{I}_{n_{\rm{t}}} -{\mathbf{K}_{k-1}}\hat{\mathbf{C}}_{:,1:k-1}^{\rm{H}}\big{)}\hat{\mathbf{h}}_{k-1|k-2}^{(n)}+{\mathbf{K}_{k-1}}{{\mathbf{y}}}_{1:k-1}^{(n)}
	\end{align}
	where $\hat{\mathbf{C}}_{:,k}$ is given in \eqref{uuubnkop}, the Kalman filter gain ${\mathbf{K}_{k-1}}$ at the $(k-1)$-th time index is given as
	\begin{align}\label{hfdghdgfgd44}
	{\mathbf{K}_{k-1}}&=R(\rho){\mathbf{\Sigma}}_{k-1|k-2}^{(n)}\hat{\mathbf{C}}_{:,1:k-1}\\ \nonumber
	&\hspace{-1em}\times \Big{(}\hat{\mathbf{C}}_{:,1:k-1}^{\rm{H}}{\mathbf{\Sigma}}_{k-1|k-2}^{(n)}\hat{\mathbf{C}}_{:,1:k-1}+\sigma_{\rm{w}}^2\mathbf{I}_{k-1}\Big{)}^{-1},
	\end{align}
	and ${\mathbf{\Sigma}}_{k|k-1}^{(n)}$ is recursively obtained as in \eqref{555mm3n3ll11}.
	The initial channel estimation, i.e., $\hat{\mathbf{h}}_{n_{\rm{p}}|n_{\rm{p}}-1}^{(n)}$, and its corresponding covariance matrix $\hat{\mathbf{\Sigma}}_{n_{\rm{p}}|n_{\rm{p}}-1}^{(n)}$ are obtained by using \eqref{03vzn3099vb41} and \eqref{7777777777774444444222222222221} for the pilot symbols in ${\mathbf{P}}$.

	By fixing the number of observations to $n_{\rm{p}}$ time index for one-step channel prediction, a simplified \ac{dd}-\ac{ce} based on Kalman filter is obtained.
	In this case, ${{\mathbf{y}}}_{1:k-1}^{(n)}$ and ${\hat{\mathbf{C}}_{n,1:k-1}}$, $n=1,2, \cdots, $ $ n_{\rm{t}}$, in \eqref{77766ggggb777uu90bee}, \eqref{hfdghdgfgd44}, and \eqref{555mm3n3ll11}
	are replaced with ${{\mathbf{y}}}_{k-n_{\rm{p}}n,k-1}^{(n)}$ and ${\hat{\mathbf{C}}_{n,k-n_{\rm{p}}:k-1}}$. Also, ${\mathbf{I}}_{k-1}$ is changed to ${\mathbf{I}}_{n_{\rm{p}}}$.

	\begin{figure*}
		\setcounter{equation}{26}		
		\begin{align}\label{7777777O0P8}
		{\mathbf{\Sigma}}_{k|k-1}^{(n)} & \hspace{-0.3em}\triangleq \hspace{-0.2em} \mathbb{E}\Big{\{}(\hat{\mathbf{g}}_{k|k-1}^{(n)}\hspace{-0.1em}-\hspace{-0.1em}{\mathbf{g}}_{k}^{(n)})(\hat{\mathbf{g}}_{k|k-1}^{(n)}\hspace{-0.1em}-\hspace{-0.1em}{\mathbf{g}}_{k}^{(n)})^{\rm{H}}\big{|}{{\mathbf{y}}}_{1:k-1}^{(n)}\Big{\}}
		={\mathbf{I}}_{n_{\rm{t}}}\hspace{-0.1em}-{{\mathbf{Q}}}_{k-1}^{\rm{H}}{\mathbf{F}}_{k-1}^{\rm{H}}\Big{(}{\mathbf{F}}_{k-1}{{\mathbf{R}}}_{1:k-1}^{\rm{d}}
		{\mathbf{F}}_{k-1}^{\rm{H}}+\sigma_{\rm{w}}^2\mathbf{I}_{k-1}\Big{)}^{-1}{\mathbf{U}}_{k-1}{{\mathbf{A}}}_{k-1}
		\end{align}
		\hrulefill
	\end{figure*}

	\subsection{\ac{dd}-\ac{ce} for \ac{stbc}}
	In this section, we obtain the \ac{dd}-\ac{ce} for \ac{stbc} transmission by using $n_{\rm{x}}$-step optimal Wiener predictor.
	
	\subsubsection{\ac{dd}-\ac{ce} based on Optimal Wiener Predictor}
	\ac{dd}-\ac{ce} for \ac{stbc} transmission is more challenging compared to the spatial multiplexing since information symbols are jointly detected based on the $n_{\rm{x}}$  observations corresponding to the transmitted \ac{stbc}.
	Hence, the optimal one-step channel prediction using the optimal Wiener predictor cannot be employed. For an \ac{stbc} code with $n_{\rm{x}}$ time interval, $n_{\rm{x}}$-step channel predictor is required.
Let us define
\setcounter{equation}{21}
	\begin{align}
	{\mathbf{g}}_{k}^{(n)}&\triangleq\Big{[}\big{(}\mathbf{h}_{k}^{(n)}\big{)}^T~\big{(}\mathbf{h}_{k+1}^{(n)}\big{)}^T~\cdots ~ \big{(}\mathbf{h}_{k+n_{\rm{x}}-1}^{(n)}\big{)}^T\Big{]}^T,
	\end{align}
	where $k=n_{\rm{p}}+1+\alpha n_{\rm{x}}$ and $\alpha=0,1,\cdots,(n_{\rm{b}}-1)$.
	
	\ac{dd}-\ac{ce} for \ac{stbc} transmission using the optimal $n_{\rm{x}}$-step Wiener predictor for the $n$-th receive antenna is expressed as
	\begin{align}\label{77766ggggb777uu90bnnmn}
	\hat{\mathbf{g}}_{k|k-1}^{(n)}=\mathbb{E}\big{\{}\mathbf{g}_{k}^{(n)}|{{\mathbf{y}}}_{1:k-1}^{(n)},{\hat{\mathbf{C}}_{:,1:k-1}}\big{\}},\,\,\,\,\ k=n_{\rm{p}}+1+\alpha n_{\rm{x}},
	\end{align}
	\begin{equation}\label{stbc_dec}
	\hat{\mathbf{C}}_{:,k:k+n_{\rm{x}}-1}=
	\begin{cases}
	T_{\rm{stbc}}\big{(}{{\mathbf{Y}}}_{k},\hat{\mathbf{G}}_{k}\big{)} &  k=n_{\rm{p}}+1+\alpha n_{\rm{x}} \\
	\mathbf{P}_{:,k}& 1 \leq k\leq n_{\rm{p}},
	\end{cases}
	\end{equation}
	where $T_{\rm{stbc}}$ is either the optimal \ac{ml} detector or a suboptimal detector, and
\begin{subequations}
	\begin{align}
	{\mathbf{Y}}_{k}&\triangleq\Big{[}{\mathbf{y}}_{k} ~{\mathbf{y}}_{k+1} ~\cdots ~ {\mathbf{y}}_{k+n_{\rm{x}}-1} \Big{]} \\
	\hat{\mathbf{G}}_{k}&\triangleq\Big{[}\hat{\mathbf{g}}_{k|k-1}^{(1)}~\hat{\mathbf{g}}_{k|k-1}^{(2)}~\cdots ~ \hat{\mathbf{g}}_{k|k-1}^{(n_{\rm{r}})}\Big{]}
	\end{align}
\end{subequations}
with ${\mathbf{y}}_{k}$ as the receiver vector at the $k$-th time index.

	For Rayleigh fading channel and \ac{awgn} at the receiver, one can write \eqref{77766ggggb777uu90bnnmn} as
	\begin{align} \label{03vzn3099vb41hhhhu}
	\hat{\mathbf{g}}_{k|k-1}^{(n)}&={\mathbf{\Sigma}}_{\mathbf{g}_{k}^{(n)},{\bf{y}}_{1:k-1}^{(n)}}\mathbf{\Sigma}_{{\bf{y}}_{1:k-1}^{(n)}}^{-1}{\bf{y}}_{1:k-1}^{(n)}
	 \\ \nonumber
	&={{\mathbf{Q}}}_{k-1}^{\rm{H}}{\mathbf{F}}_{k-1}^{\rm{H}} \Big{(}{\mathbf{F}}_{k-1}{{\mathbf{R}}}_{k-1}^{\rm{d}}
	{\mathbf{F}}_{k-1}^{\rm{H}}+\sigma_{\rm{w}}^2\mathbf{I}\Big{)}^{-1}{\mathbf{y}}_{1:k-1}^{(n)},
	\end{align}
	where
	${\mathbf{\Sigma}}_{\mathbf{g}_{k}^{(n)},{\bf{y}}_{1:k-1}^{(n)}}  \hspace{-0.2em}\triangleq \hspace{-0.2em} \mathbb{E} \big{\{}{\mathbf{g}}_{k}^{(n)}{{\bf{y}}_{1:k-1}^{(n){\rm{H}}}}\big{\}}$,
	${\mathbf{F}}_{k-1}\hspace{-0.2em}\triangleq \hspace{-0.2em}$
	${\mathbf{I}}_{n_{\rm{x}}}\otimes {{\mathbf{U}}}_{k-1}$,
	${{\mathbf{Q}}}_{k-1}={\rm{bdiag}}\Big{(}{\rm{diag}}\big{(}{\mathbf{1}}_{n_{\rm{t}}}\otimes{\mathbf{r}}(1,k-1)\big{)}, $ ${\rm{diag}}
	\big{(}{\mathbf{1}}_{n_{\rm{t}}}\otimes{\mathbf{r}}(2,k)\big{)},$ $\cdots ~ , {\rm{diag}}\big{(}{\mathbf{1}}_{n_{\rm{t}}}\otimes{\mathbf{r}}(n_{\rm{x}},k+n_{\rm{x}}-2)\big{)}\Big{)}$,
	with ${\mathbf{r}}{(u,v)}$ as in \eqref{vvooo90}.

The \ac{mmse} of the optimal $n_{\rm{x}}$-step channel predictor for \ac{stbc} transmission at the $k$-th time index is given in \eqref{7777777O0P8}.
   Similar to spatial multiplexing transmission, a Weiner filter of order $n_{\rm{p}}$ can be used for
	$n_{\rm{x}}$-step channel prediction in \ac{stbc} transmission to reduce the computational complexity.

\section{Deep Learning for Channel Estimation}\label{sec:DL_CE}
	The main idea behind the proposed \ac{dl}-based \ac{dd}-\ac{ce} is to employ a trained \ac{dnn} as channel predictor
	to remove the need for channel statistics estimation, such as the exact Doppler spread, $\rho$, which is a challenging task especially in highly dynamic vehicular environments.
	Considering the substantial capability of \ac{dl} in learning nonlinear functions, a single \ac{dnn} can make a channel prediction for a wide range of Doppler rates for highly dynamic vehicular channels.
	The proposed \ac{dl}-based predictor is efficient in many vehicular channels even the those without an explicit mathematical model, where the optimal Weiner filter and \ac{kf} channel predictors are not applicable.
	
	The proposed \ac{dl}-based \ac{dd}-\ac{ce} algorithm is composed of an estimation step and a decoding step at each time index.
The estimation step consists of two stages: prediction and update. The prediction stage predicts the channel forward from measurement time.
For spatial multiplexing one-step channel prediction and for \ac{stbc} $n_{\rm{x}}$-step channel prediction is required prior to decoding.
The update stage is followed by the decoding step, and it uses the decoded \ac{stbc} and latest measurement to modify the channel prediction through a \ac{r}-\ac{mmse} algorithm.
In our \ac{dd}-\ac{ce} algorithm, the prediction stage of channel estimation is implemented through a \ac{dnn}. In the decoding step, joint \ac{ml} decoding  of the information symbols is performed.
In the following subsections, first we present the design of the \ac{dnn} $k$-step predictor and then we propose our algorithm.

	\subsection{Channel Prediction Using \ac{dl}}
   In the \ac{dl}-based channel prediction, we estimate future channel coefficients using past estimates.
   This is different from Bayesin tracking solutions, such as Winer filter and \ac{kf}, where predictions are made based on previous observations.
   Channel prediction based on all previous estimates (similar to the optimal Winer filter)
   is highly costly in terms of computational complexity, especially for the latter symbols.
	Moreover, such a design requires a time-varying \ac{dnn} with increasing input layer size as the \ac{dd}-\ac{ce} algorithm runs from one time index to the next.
	To avoid these challenges and simplify the \ac{dnn}, only the $n_{\rm{p}}$ previous estimated channel coefficients are involved in one-step channel prediction for spatial multiplexing and
   $n_{\rm{x}}$-step channel prediction for \ac{stbc}s.

Since channel prediction in our algorithm is based on $n_{\rm{p}}$ previous estimated channel coefficients, we can train the \ac{dnn} for true channel realizations in the training phase.
In practice, for fading channels without concrete mathematical model, the true values of \ac{mimo} channels can be obtained through the transmission of pilot symbols with value one.

For the prediction stage, two different \ac{dnn}s are trained to independently predict the real and imaginary parts of the \ac{mimo} fading channels.

Let us consider the $j$-th, $j=1,2,\cdots,N_{\rm{t}}$,
 training sample vector
	\setcounter{equation}{27}		
\begin{align}
{\tilde{\mathbf{h}}}{[j]}\triangleq{\mathbf{x}}{[j]}+i{\mathbf{z}}{[j]}=\Re\big{\{}\tilde{\mathbf{h}}{[j]}\big{\}}+i\Im\big{\{}\tilde{\mathbf{h}}{[j]}\big{\}},
\end{align}
    \begin{equation}
    \tilde{\mathbf{h}}{[j]}\triangleq
    \begin{bmatrix}
\bar{\mathbf{h}}_{1}[j] ~ \bar{\mathbf{h}}_{2}[j] ~ \cdots ~
\bar{\mathbf{h}}_{n_{\rm{x}}+n_{\rm{p}}}[j]
\end{bmatrix},
   \end{equation}
    \begin{equation}
    \bar{\bf{h}}_k{[j]} \triangleq
    \begin{bmatrix}
\big{(}\mathbf{h}_{k}^{(1)}[j]\big{)}^T ~ \big{(}\mathbf{h}_{k}^{(2)}[j]\big{)}^T ~ \cdots ~ \big{(}\mathbf{h}_{k}^{(n_{\rm{r}})}[j]\big{)}^T
\end{bmatrix}\\
   \end{equation}
where $\mathbf{h}_{k}^{(n)}[j]=\begin{bmatrix}
	h_{n1,k}[j] & h_{n2,k}[j] & \dots  & h_{nn_{\rm{t}},k}[j]
	\end{bmatrix}^T$ is the complex-valued fading channel coefficients between the $n$-th receive antenna and all $n_{\rm{t}}$ transmit antennas at the $k$-th time index of the $j$-th training sample.
The $N_{\rm{t}}$ training sample vectors are independently generated, and Doppler spread, $\rho$, associated with each training vector is uniformly distributed in $[{\rho}_{\rm{min}} , {\rho}_{\rm{max}}]$.
The first $u\triangleq
n_{\rm{t}}n_{\rm{r}}n_{\rm{p}}$ entries of each training vector, i.e., $\tilde{\bf{h}}_{1:u}{[j]}$ are used as the input of the \ac{dnn}s.
Our target is to train the \ac{dnn} to produce the desired output vector i.e., $\tilde{\bf{h}}_{u+1:v}{[j]}$, $v\triangleq
n_{\rm{t}}n_{\rm{r}}(n_{\rm{x}}+n_{\rm{p}})$, which is equivalent to $n_{\rm{x}}$-step channel prediction.

	During the training phase, the \ac{dnn}s learn two nonlinear transformations, $\Psi_{\rm{r}}: \mathbb{R}^{u} \rightarrow \mathbb{R}^{v}$ and $\Psi_{\rm{I}}: \mathbb{R}^{u} \rightarrow \mathbb{R}^{v}$, which maps the input vector ${\bf{x}}_{1:u}{[j]}$ to ${\bf{x}}_{u+1:v}{[j]}$ and ${\bf{z}}_{1:u}{[j]}$ to ${\bf{z}}_{u+1:v}{[j]}$ as
	\begin{subequations}\label{eq:dnn}
    \begin{align}
    	{\bf{x}}_{u+1:v}{[j]} =\Psi_{\rm{r}}({\bf{x}}_{1:u}{[j]};{\bf{\Theta}}_1)\\
    	{\bf{z}}_{u+1:v}{[j]} =\Psi_{\rm{I}}({\bf{z}}_{1:u}{[j]};{\bf{\Theta}}_2),
    \end{align}
	\end{subequations}
where $\boldsymbol{\Theta}_1$ and $\boldsymbol{\Theta}_2$ are the set of the \ac{dnn}s parameters.
	These parameters are obtained by minimizing the following \ac{ls} loss function in the off-line training phase.
	\begin{equation}\label{eq:lossFunction}
	\text{Loss}{({\bf{\Theta}}_i)}=\frac{1}{N_{\rm{t}}}\sum_{j=1}^{N_{\rm{t}}} \Big\| {\bf{x}}_{u+1:v}{[j]}- \Psi({\bf{x}}_{1:u}{[j]};{\bf{\Theta}}_i)~ \Big\|^2,\,\,\,\ i=1,2.
	\end{equation}
As seen, channel prediction is formulated as a regression task to estimate parameter vector ${\bf{\Theta}}_i$, $i=1,2$, given the training data set $\big{(}{\bf{x}}_{1:u}{[j]},\tilde{\bf{x}}_{u+1:v}{[j]}\big{)}$, and $\big{(}{\bf{z}}_{1:u}{[j]},\tilde{\bf{z}}_{u+1:v}{[j]}\big{)}$, $j=1,2,\cdots,N_{\rm{t}}$.

	Designing a \ac{dnn} with an appropriate layered structure yields an accurate predictor functions in \eqref{eq:dnn}. This is crucial for precise channel prediction when the exact value of Doppler rate is unknown.
	In particular, the number of hidden layers and the number of neurons in each layer affect the range of Doppler rate that can be supported by the \ac{dnn}.
	Our simulation experiments based on existing guidelines for neural network architecture selection show that a \ac{dnn} with the layered structure in Tables \ref{8889755rruop} and \ref{999773bbcvvxx} results in accurate channel prediction for Almauti and Tarokh \ac{stbc}s in \cite{alamouti} and \cite{tarokh2} for the range of Doppler rate $[\rho_{\rm{min}}, \rho_{\rm{max}}]$, where $0.001\le\rho_{\rm{max}}-\rho_{\rm{min}}\le0.1$, $\rho_{\rm{min}} \ge 0$ and $\rho_{\rm{max}} \le 0.1$.
		\begin{table}[t!]
		\centering
		\caption{List of DNN layers and outputs}
		\label{table}
		\begin{tabular}{l*{6}{c}r}
			Name              & Output Dimensions  \\
			\hline
			Sequence Input & $n_t \times n_r \times n_p$   \\
			Dense + CReLU  ($1^{st}$)       & 128  \\
			Dense + CReLU  ($2^{nd}$)     & 128  \\
			Regression Output    & $n_t \times n_r \times n_x$
		\end{tabular}\label{8889755rruop}
	\end{table}
	\begin{table}[t!]
		\centering
		\caption{List of DNN functions}
		\label{table2}
		\begin{tabular}{l*{6}{c}r}
			Name              & Function  \\
			\hline
			CReLU & $f(a)=au(a)+(a-1)u(a-1)$   \\
			RMSE         & $l(u,\hat{u})=\vert|u-\hat{u}|\vert^2_2$
		\end{tabular}\label{999773bbcvvxx}
	\end{table}
	
	\subsection{\ac{dl}-Based \ac{dd}-\ac{ce} Algorithm}
Let us stack the channel coefficients of the fading channels over the transmission packet as an ${n_{\rm{t}}}{n_{\rm{r}}(n_{\rm{b}}n_{\rm{x}}+n_{\rm{p}}})\times 1$ dimensional vector
\begin{align}
\tilde{\mathbf{h}}\triangleq
\begin{bmatrix}
\bar{\mathbf{h}}_{1} ~ \bar{\mathbf{h}}_{2}~ \cdots ~
\bar{\mathbf{h}}_{n_{\rm{x}}n_{\rm{b}}+n_{\rm{p}}}
\end{bmatrix}^T,
\end{align}
where
    \begin{equation}
    \bar{\bf{h}}_k \triangleq
    \begin{bmatrix}
\big{(}\mathbf{h}_{k}^{(1)}\big{)}^T ~ \big{(}\mathbf{h}_{k}^{(2)}\big{)}^T ~ \cdots ~ \big{(}\mathbf{h}_{k}^{(n_{\rm{r}})}\big{)}^T
\end{bmatrix},
   \end{equation}
and
\begin{align}
\mathbf{h}_{k}^{(n)}=\begin{bmatrix}
	h_{n1,k} & h_{n2,k} & \dots  & h_{nn_{\rm{t}},k}
	\end{bmatrix}^T.
\end{align}
Using the proposed \ac{dl}-based $n_{\rm{x}}$-step channel predictor, we can design a \ac{dd}-\ac{ce} without knowledge of exact Doppler rate value.
For each \ac{stbc}, the corresponding $n_{\rm{x}}n_{\rm{t}}n_{\rm{r}}$ channel coefficients are predicted based on the previously predicted and updated $n_{\rm{p}}n_{\rm{t}}n_{\rm{r}}$ channel coefficients.

By employing the learned predictor functions $\Psi_{\rm{r}}$ and $\Psi_{\rm{i}}$,
channel prediction for the $k$-th \ac{stbc} in the packet is expressed as
\begin{align}\label{dsdsdsqq}
&\hspace{-3em}\hat{{\bf{x}}}_{n_{\rm{t}}n_{\rm{r}}((k-1)n_{\rm{x}}+n_{\rm{p}})+1:n_{\rm{t}}n_{\rm{r}}(kn_{\rm{x}}+n_{\rm{p}})}^{\rm{p}} \\ \nonumber
&=\Psi_{\rm{r}}\Big{(}\hat{\bf{x}}_{n_{\rm{t}}n_{\rm{r}}(k-1)n_{\rm{x}}+1:n_{\rm{t}}n_{\rm{r}}((k-1)n_{\rm{x}}+n_{\rm{p}})}^{\rm{u}};{\bf{\Theta}}\Big{)},
\end{align}
\begin{align}\label{dsdsdsqqrererer3}
&\hspace{-3em}\hat{{\bf{z}}}_{n_{\rm{t}}n_{\rm{r}}((k-1)n_{\rm{x}}+n_{\rm{p}})+1:n_{\rm{t}}n_{\rm{r}}(kn_{\rm{x}}+n_{\rm{p}})}^{\rm{p}}  \\ \nonumber
&=\Psi_{\rm{i}}\Big{(}\hat{\bf{z}}_{n_{\rm{t}}n_{\rm{r}}(k-1)n_{\rm{x}}+1:n_{\rm{t}}n_{\rm{r}}((k-1)n_{\rm{x}}+n_{\rm{p}})}^{\rm{u}};{\bf{\Theta}}\Big{)},
\end{align}
\begin{align}\label{99979RRRP7NXXZWEOCG}
&\hspace{-3em}\hat{\bf{h}}_{n_{\rm{t}}n_{\rm{r}}((k-1)n_{\rm{x}}+n_{\rm{p}})+1:n_{\rm{t}}n_{\rm{r}}(kn_{\rm{x}}+n_{\rm{p}})}^{\rm{p}}\\ \nonumber
&=\hat{\bf{x}}_{n_{\rm{t}}n_{\rm{r}}((k-1)n_{\rm{x}}+n_{\rm{p}})+1:n_{\rm{t}}n_{\rm{r}}(kn_{\rm{x}}+n_{\rm{p}})}^{\rm{p}} \\ \nonumber
&+i\hat{\bf{z}}_{n_{\rm{t}}n_{\rm{r}}((k-1)n_{\rm{x}}+n_{\rm{p}})+1:n_{\rm{t}}n_{\rm{r}}(kn_{\rm{x}}+n_{\rm{p}})}^{\rm{p}}.
\end{align}
where $\hat{\bf{x}}_.^{\rm{u}}$ and $\hat{\bf{z}}_.^{\rm{u}}$ are real and imaginary parts of the channel coefficients after \ac{r}-\ac{mmse} modification based on the decocted \ac{stbc} and latest measurement in the update step which will be explained in the following.

After channel prediction stage, the predicted channel coefficients in \eqref{99979RRRP7NXXZWEOCG} are used for decoding. Decoding can be implemented through optimal or suboptimal algorithms.

We consider a decoding algorithm $T_{\rm{stbc}}$ (details on the decoding is provided in the next subsection) and write the decoded $k$-th \ac{stbc} as
\begin{align}\label{6iuhqhd901213123}
\hat{{{\bf{C}}}}_{:,n_{\rm{p}}+(k-1)n_{\rm{x}}+1:n_{\rm{p}}+kn_{\rm{x}}} &\\ \nonumber
&\hspace{-8em}=T_{\rm{stbc}}\big{(}\hat{\bf{h}}_{n_{\rm{t}}n_{\rm{r}}((k-1)n_{\rm{x}}+n_{\rm{p}})+1:n_{\rm{t}}n_{\rm{r}}(kn_{\rm{x}}+n_{\rm{p}})}^{\rm{p}},\tilde{\mathbf{y}}_k\big{)},
\end{align}
where ${{{\bf{C}}}}$ is given in \eqref{packet}, and $\tilde{\mathbf{y}}_k$ is the observation vector associated with the $k$-th \ac{stbc} given as
\begin{align}
\tilde{\mathbf{y}}_k=
	\begin{bmatrix}
		{\mathbf{y}}_{n_{\rm{p}}+(k-1)n_{\rm{x}}+1}\\
		{\mathbf{y}}_{n_{\rm{p}}+(k-1)n_{\rm{x}}+2} \\
		\vdots \\
	    {\mathbf{y}}_{n_{\rm{p}}+kn_{\rm{x}}}
		\end{bmatrix}.
\end{align}
In the update stage of the estimation step, the input of the \ac{dnn}s for the next prediction are updated using \ac{r}-\ac{mmse} algorithm.
The \ac{r}-\ac{mmse} algorithm exploits the decoded \ac{stbc} $\hat{{{\bf{C}}}}_{:,n_{\rm{p}}+(k-1)n_{\rm{x}}+1:n_{\rm{p}}+kn_{\rm{x}}}$ in \eqref{6iuhqhd901213123}
and previously decoded \ac{stbc}s or preambles $\hat{{{\bf{C}}}}_{:,(k-1)n_{\rm{x}}+1:n_{\rm{p}}+(k-1)n_{\rm{x}}}$ to update the input of the \ac{dnn}s.

Let us write the observation vector associated with the \ac{stbc}s or preambles $ \hat{{{\bf{C}}}}_{:,kn_{\rm{x}}+1:n_{\rm{p}}+kn_{\rm{x}}}$ as
\begin{align}\label{7yuuueete2364347}
\tilde{\mathbf{y}}_k^{\rm{u}}=\mathbf{E}_k^{\rm{u}}\mathbf{\Upsilon}_k^{\rm{u}}+\mathbf{w}_k^{\rm{u}},
\end{align}
where
\begin{align}
\tilde{\mathbf{y}}_k^{{\rm{u}}}\triangleq
	\begin{bmatrix}
		{\mathbf{y}}_{kn_{\rm{x}}+1}\\
		{\mathbf{y}}_{kn_{\rm{x}}+2} \\
		\vdots \\
	    {\mathbf{y}}_{kn_{\rm{x}}+n_{\rm{p}}}
		\end{bmatrix}.
\end{align}
\begin{align}
\mathbf{E}_k^{\rm{u}}\triangleq {\rm{bdiag}}\Big{(} ~\mathbf{E}(k,1)~\mathbf{E}(k,2)~ \cdots ~\mathbf{E}(k,n_{\rm{p}})\Big{)}
\end{align}
\begin{align}
\mathbf{E}(k,m) \triangleq {\mathbf{I}}_{n_{\rm{t}}n_{\rm{r}}}\otimes \hat{\mathbf{C}}_{:,kn_{\rm{x}}+m}^T,
\end{align}
\begin{align}
\mathbf{\Upsilon}_k^{\rm{u}}\triangleq \hat{{\bf{h}}}_{kn_{\rm{t}}n_{\rm{r}}n_{\rm{x}}+1:n_{\rm{t}}n_{\rm{r}}(kn_{\rm{x}}+n_{\rm{p}})}^{\rm{u}},
\end{align}
\begin{align}
\mathbf{w}_k^{\rm{u}}\triangleq
\begin{bmatrix}
\mathbf{w}(k,1) & \mathbf{w}(k,2)  & \mathbf{w}(k,n_{\rm{p}})                        \end{bmatrix}^T
\end{align}
and
\begin{equation}
\mathbf{w}(k,m)\triangleq
\begin{bmatrix}
w^{(1)}_{kn_{\rm{x}}+m} & w^{(2)}_{kn_{\rm{x}}+m}, \cdots w^{(n_{\rm{r}})}_{kn_{\rm{x}}+m}                             \end{bmatrix}.
\end{equation}

The \ac{r}-\ac{mmse} replaces the true value of the Doppler spread in the covariance matrix used in the \ac{mmse} estimator with the average Doppler spreads as
\begin{align}
\bar{\rho}=\frac{\rho_{\rm{max}}+\rho_{\rm{min}}}{2}.
\end{align}
Hence, the doppler rate $\rho$ in the covariance matrix ${\mathbf{R}}_{n_{\rm{t}}n_{\rm{p}}-1}$ in \eqref{corrttt} is replaced with $\bar{\rho}$ and then ${\mathbf{\Omega}} \triangleq   {\mathbf{I}}_{n_{\rm{r}}}\otimes {\mathbf{R}}_{n_{\rm{t}}n_{\rm{p}}-1}$ is used to obtain the updated channel coefficients as
\begin{align}\label{eq:rmmse}
\hat{{\bf{h}}}_{kn_{\rm{t}}n_{\rm{r}}n_{\rm{x}}+1:n_{\rm{t}}n_{\rm{r}}(kn_{\rm{x}}+n_{\rm{p}})}^{\rm{u}}
=
{\mathbf{\Omega}}\big{(}{\mathbf{E}}_{k}^{{\rm{u}}}\big{)}^{\rm{H}}\big{(}{\mathbf{E}}_{k}^{\rm{u}}
{\mathbf{\Omega}}\big{(}{\mathbf{E}}_{k}^{{\rm{u}}}\big{)}^{\rm{H}}+\sigma_{\rm{w}}^2\mathbf{I}\big{)}^{-1}\tilde{\mathbf{y}}_k^{\rm{u}}.
\end{align}


\subsection{\ac{ml} Decoding Algorithm for STBC Design}
Let us write the received vector associated with the $k$-th \ac{stbc} in the packet as
\begin{align}\label{7yuuueete23643472}
\tilde{\mathbf{y}}_k=\mathbf{E}_k^{\rm{p}}\mathbf{\Upsilon}_k^{\rm{p}}+\mathbf{w}_k^{\rm{p}}
\end{align}
where $\mathbf{\Upsilon}_k^{\rm{p}}\triangleq \hat{{\bf{h}}}_{n_{\rm{t}}n_{\rm{r}}((k-1)n_{\rm{x}}+n_{\rm{p}})+1:n_{\rm{t}}n_{\rm{r}}(kn_{\rm{x}}+n_{\rm{p}})}^{\rm{p}}$,
$\mathbf{E}_k^{\rm{p}}\triangleq {\rm{bdiag}}\Big{(} ~\mathbf{X}(k,1)~\mathbf{X}(k,2)~ \cdots ~\mathbf{X}(k,n_{\rm{x}})\Big{)}$, where
$\mathbf{X}(k,m) \triangleq {\mathbf{I}}_{n_{\rm{t}}n_{\rm{r}}}\otimes \hat{\mathbf{C}}_{:,(k-1)n_{\rm{x}}+m}^T$.

By using \eqref{7yuuueete23643472}, the \ac{ml} decoding of the information symols in the $k$-th \ac{stbc} is obtained as
	\begin{equation} \label{eq:ml}
	\hat{\mathbf{s}}_{k} = \arg\max_{s_1,\cdots,s_N \in \mathbb{D}} f(\tilde{\mathbf{y}}_k|\mathbf{s}_{k},\mathbf{\Upsilon}_k^{\rm{p}}).
	\end{equation}
For \ac{awgn} noise, one can easily write
    \begin{equation}\label{eq:mlg}
    	\hat{\mathbf{s}}_{k}  = \arg\max_{s_1,\cdots,s_N \in \mathbb{D}} \frac{e^{\tilde{\mathbf{y}}_k^{\rm{ H}}{\mathbf{\Gamma}} ^{-1}\tilde{\mathbf{y}}_k}}{|\pi{\mathbf{\Gamma}} |},
    \end{equation}
where
	\begin{equation*}
		{\mathbf{\Gamma}}  = \mathbb{E}\{\tilde{\mathbf{y}}_k\tilde{\mathbf{y}}_k^{\rm{H}}\} =\mathbf{E}_k^{\rm{p}}\mathbf{\Upsilon}_k\mathbf{\Upsilon}_k^{\rm{H}}\big{(}\mathbf{E}_k^{\rm{p}}\big{)}^{\rm{H}}+\sigma_{\rm{w}}^2\mathbf{I}_{n_{\rm{r}}n_{\rm{x}}}.
	\end{equation*}
	and after some mathematical manipulations, it results in
	\begin{equation}\label{yytrrom}
		\hat{\mathbf{s}}_{k}= \arg\max_{s_1,\cdots,s_N \in \mathbb{D}} (\tilde{\mathbf{y}}_k)^{\rm{H}}{\mathbf{\Gamma}} ^{-1}(\tilde{\mathbf{y}}_k)+\ln |{\mathbf{\Gamma}} |.
	\end{equation}
There is no further simplification for the detection problem in \eqref{yytrrom}; hence,  it should be solved through exhaustive search or dynamic programming.
\subsubsection{Alamouti Decoding}
For Alamouti \ac{stbc}, the decoding in \eqref{yytrrom} can be formulated as an \ac{ls} optimization problem.

Let us write the received vector associated with the $k$-th \ac{stbc} as
\begin{align}\label{7yuuueete23643471}
\breve{\mathbf{y}}_k=\mathbf{B}_k\mathbf{s}_k+\mathbf{w}_k
\end{align}
where $\breve{\mathbf{y}}_k \triangleq
	\begin{bmatrix}
{\mathbf{y}}_{n_{\rm{p}}+2k-1}^T & {\mathbf{y}}_{{n_{\rm{p}}+2k}}^{\rm{H}}
	\end{bmatrix}^T$,

	\begin{equation} \label{eq:ch}
		\mathbf{B}_{k} \triangleq
		\begin{bmatrix}
			{\mathbf{v}}(k,1) & {\mathbf{v}}(k,3) \\
			{\mathbf{v}}(k,5)^* & -{\mathbf{v}}(k,7)^*
		\end{bmatrix},
	\end{equation}
and ${\mathbf{v}}(k,m)\triangleq \hat{{\bf{h}}}_{4(2(k-1)+n_{\rm{p}})+m:4(2(k-1)+n_{\rm{p}})+m+1}$. One can easily show that the \ac{ml} decoding based on the observation model in \eqref{7yuuueete23643471} leads to the following \ac{ls} optimization.
	\begin{align}\label{eq:mldetection}
	\hat{\mathbf{s}}_t=
	\underset{{s_{1}} , {s_{2}} \in \mathbb{D}} {{\rm{arg}} \min} \,\,\,\ \Big{\Vert}\breve{\mathbf{y}}_k-\mathbf{B}_{k}\mathbf{s}_{k}\Big{\Vert}^2.
	\end{align}
	
	The procedure of our \ac{dl}-based algorithm is briefly presented in Algorithm \ref{alg}.

\begin{algorithm}[t!]
	\caption{\ac{dl}-based \ac{dd}-\ac{ce} with \ac{ml} Decoding Algorithm for STBC Design}
	\begin{algorithmic}[1]
		\Statex \textbf{Input:} $\tilde{\mathbf{y}}_1, ~ \cdots,~ \tilde{\mathbf{y}}_{n_{\rm{b}}}$, $\Psi_{\rm{r}}$ and $\Psi_{\rm{i}}$
		\Statex \textbf{Output:} $\hat{\mathbf{h}}_{1}, ~ \hat{\mathbf{h}}_{2},~ \cdots ~, \hat{\mathbf{h}}_{n_{\rm{b}}}$
		\State \textbf{for} $i:=+1$~to~$n_{\rm{b}}$ \textbf{do}
		\State \textit{Prediction step:}
		\Statex Stack real and imaginary parts of previous $n_{\rm{p}}$ channels to $\Psi_{\rm{r}}$ and $\Psi_{\rm{i}}$, respectively to obtain the channels of the $i$-th STBC block as \eqref{dsdsdsqq} and \eqref{dsdsdsqqrererer3}.
		\State \textit{Decoding step:}
		\Statex Use the derived \ac{ml} Decoding Algorithm in \eqref{yytrrom} and predicted channels to detect the $i$-th transmitted STBC block $\hat{{{\bf{C}}}}_{:,n_{\rm{p}}+(i-1)n_{\rm{x}}+1:n_{\rm{p}}+in_{\rm{x}}}$.
		\State \textit{Updating step:}
		\Statex By employing the detected STBC block, update the predicted channels by \ac{r}-\ac{mmse} as follows in \eqref{eq:rmmse} to obtain $\hat{{\bf{h}}}_{in_{\rm{t}}n_{\rm{r}}n_{\rm{x}}+1:n_{\rm{t}}n_{\rm{r}}(in_{\rm{x}}+n_{\rm{p}})}^{\rm{u}}$ as the input of the \ac{dnn} for the next prediction.
		\State \textbf{end}
	\end{algorithmic} \label{alg}
\end{algorithm}

	\section{Complexity Analysis}\label{sec:comp}
In this section we compare the computational complexity of our proposed \ac{dl}-based algorithm with \ac{mmse} \ac{dd}-\ac{ce}, first-order autoregression AR(1) \ac{dd}-\ac{ce}.

Table \eqref{8889755rruopbbbbbbbt} compares the number of floating-point
operation (real addition, substration, and multiplication) in the proposed \ac{dl}-based $n_{\rm{x}}$-step channel predictor with the Winer, CC, and AR(1) predictors.
As seen, the proposed channel predictor exhibits a lower computational complexity compared to the optimal Winer predictor of order $n_{\rm{p}}$. Moreover, compared to the DD-AR1 \cite{Gia_method} and DD-AR1 \cite{tarokh_cestbc} predictors, the proposed algorithm shows a higher computational complexity at the expense of lower \ac{ber} and propagation error.
\begin{table*}[t!]
	\centering
	\caption{Complexity Comparison between different channel predictors in \ac{dd}-\ac{ce}.}
	\label{table}
	\begin{tabular}{l*{6}{c}r}
		Name               & Number of Flops  \\
		\hline
		Wiener of order $n_{\rm{p}}$ & $n_{\rm{r}}n_{\rm{x}}(\gamma+3\gamma^3+5\gamma^2+4(n_{\rm{p}}-1)nt+6n_{\rm{t}}(n_{\rm{p}}-1)^3+4n_{\rm{t}}(n_{\rm{p}}-1)^2-2n_{\rm{t}}(n_{\rm{p}}-1))$, \,\,\,\ $\gamma \triangleq (n_{\rm{p}}-1)^2(6n_{\rm{t}}-2)+(n_{\rm{p}}-1)$\\
		DD-CC &  $n_{\rm{p}}(3n_{\rm{p}} + 2n_{\rm{p}}n_{\rm{t}} - 2n_{\rm{r}}n_{\rm{t}} + 4n_{\rm{p}}n_{\rm{t}}^2
		+ 6n_{\rm{p}}^2n_{\rm{t}}+ 3n_{\rm{p}}^2 + 6n_{\rm{p}}n_{\rm{r}}n_{\rm{t}} + 1)$\\
		DD-AR1 & $n_{\rm{p}}(3n_{\rm{p}} + 2n_{\rm{p}}n_{\rm{t}} - 2n_{\rm{r}}n_{\rm{t}} + 4n_{\rm{p}}n_{\rm{t}}^2 + 6n_{\rm{p}}^2n_{\rm{t}}+ 3n_{\rm{p}}^2 + 6n_{\rm{p}}n_{\rm{r}}n_{\rm{x}} + n_{\rm{r}}n_{\rm{t}}n_{\rm{t}}+1)$ \\
	    DL-DD  & $n_{\rm{p}}(3n_{\rm{p}} + 2n_{\rm{p}}n_{\rm{t}} - 2n_{\rm{r}}n_{\rm{t}} + 4n_{\rm{p}}n_{\rm{t}}^2
+ 6n_{\rm{p}}^2n_{\rm{t}}+ 3n_{\rm{p}}^2 + 6n_{\rm{p}}n_{\rm{r}}n_{\rm{p}} + 1)+512(n_{\rm{t}}n_{\rm{r}}(n_{\rm{x}}+n_{\rm{p}})+128)$.
	\end{tabular}\label{8889755rruopbbbbbbbt}
\vspace{-.5cm}
\hrulefill
	\end{table*}

	\section{Simulations And Results}\label{sec:sim}
	In this section we provide some performance measures to compare our proposed \ac{dl}-based \ac{dd}-\ac{ce} for \ac{mimo} communication systems with the \ac{dd}-\ac{ce} method which model channel based on first order autoregressive model in \cite{Gia_method} and the \ac{mmse} \ac{dd}-\ac{ce} provided in \cite{tarokh_cestbc} where channel is assumed to be coherent for each of \ac{stbc} block transmission.
	We denote our method by DL-DD and the methods in \cite{Gia_method} and \cite{tarokh_cestbc} by DD-AR1 and DD-CC, respectively.
	\subsection{Simulation Setup}
Unless otherwise mentioned, 
	we consider 4-QAM constellation in \ac{mimo} time-varying fading channel and run our simulations for both Rayleigh and Rician fading channels.
	We model the fading channels by Jake's Doppler spectrum, where
	the autocorrelation function of the channel is given as
	\begin{align}
	\mathbb{E}\{h_{nm,k_1}h_{nm,k_2}^*\}=&\frac{K}{K+1}e^{(-j2\pi f_{\rm{D}}cos(\alpha_0))} \\ \nonumber 
&\hspace{-3em}+\frac{\sigma_{\rm{h}}^2}{K+1}J_0(2\pi \rho(k_1-k_2))\,\,\,\,\,\,\ \rho \in [\rho_{\rm{min}} \ \rho_{\rm{max}}],\,\,\,\,\,\
	\end{align}
	with $K$ being $K$-factor, $f_{\rm{D}}$ being \ac{los} component of fading, $\sigma_{\rm{h}}^2$ being the average \ac{nlos} power of $h_{nm}$, and $\rho$ being the Doppler rate.
	Without loss of generality, we assume that the only available knowledge in the receiver side is the range of Doppler rate and not the exact value which is accessible by current channel estimators.
	The range of Doppler rate is set such that $0.001\le\rho_{\rm{max}}-\rho_{\rm{min}}\le0.1$.
	
	We provide performance measures for three different \ac{stbc}s including Alamouti STBC \cite{alamouti} which gives a rate one by $n_{\rm{t}}=2$ transmit antennas as
	\begin{equation}\label{eq:alamouti}
	{\mathbf{C}}_{\rm{Al}}^T=
	\begin{bmatrix}
	s_{1} & s_{2} \\
	-s_{2}^* & s_{1}^*
	\end{bmatrix},
	\end{equation}
Tarokh \textit{et. al}'s \ac{stbc} \cite{tarokh2} which achieves code rate $3/4$ with $n_{\rm{t}}=3$ transmit antennas given by
\begin{equation}\label{eq:tarokh}
{\mathbf{C}}_{\rm{Ta}}^T=
\begin{bmatrix}
s_{1} & s_{2} & \frac{s_{3}}{\sqrt{2}} \\
-s_{2}^* & s_{1}^* & \frac{s_{3}}{\sqrt{2}} \\
\frac{s_{3}^*}{\sqrt{2}} & \frac{s_{3}^*}{\sqrt{2}} & \frac{-s_{1}-s_{1}^*+s_{2}-s_{2}^*}{2} \\
\frac{s_{3}^*}{\sqrt{2}} & \frac{-s_{3}^*}{\sqrt{2}} & \frac{s_{2}+s_{2}^*+s_{1}-s_{1}^*}{2}
\end{bmatrix}.
\end{equation}
and the following \ac{stbc} code with code rate $3/4$ with $n_{\rm{t}}=3$, $n_{\rm{r}}=2$, and $n_{\rm{x}}=4$ as
\begin{equation}\label{eq:tarokh343435}
{\mathbf{C}}_{3/4}^T=
\begin{bmatrix}
s_{1} & s_{2} & s_{3} \\
-s_{2}^* & s_{1}^* & 0 \\
{s_{3}^*}& 0 & s_{1}^*\\
0 & {-s_{3}^*}& {s_{2}^*}
\end{bmatrix}.
\end{equation}

	The additive noise is modeled as circular symmetric zero-mean complex-valued Gaussian random variable with variance $\sigma_{\rm{w}}^2$, i.e. $w_k\sim\mathcal{CN}(0,\sigma_{\rm{w}}^2)$.
	The \ac{snr} in dB is defined as $\gamma=10\log(\sigma_{\rm{s}}^2/\sigma_{\rm{w}}^2)$, where $\sigma_{\rm{s}}^2$ is the average transmitted power.
	Unless otherwise mentioned, the length of the transmitted packet is $L=100$ and the length of the pilot is $n_{\rm{p}}=10$ and also $\sigma_{\rm{h}}^2=1$.

	\begin{figure}[]

		\centering
		\includegraphics[width=0.5\textwidth]{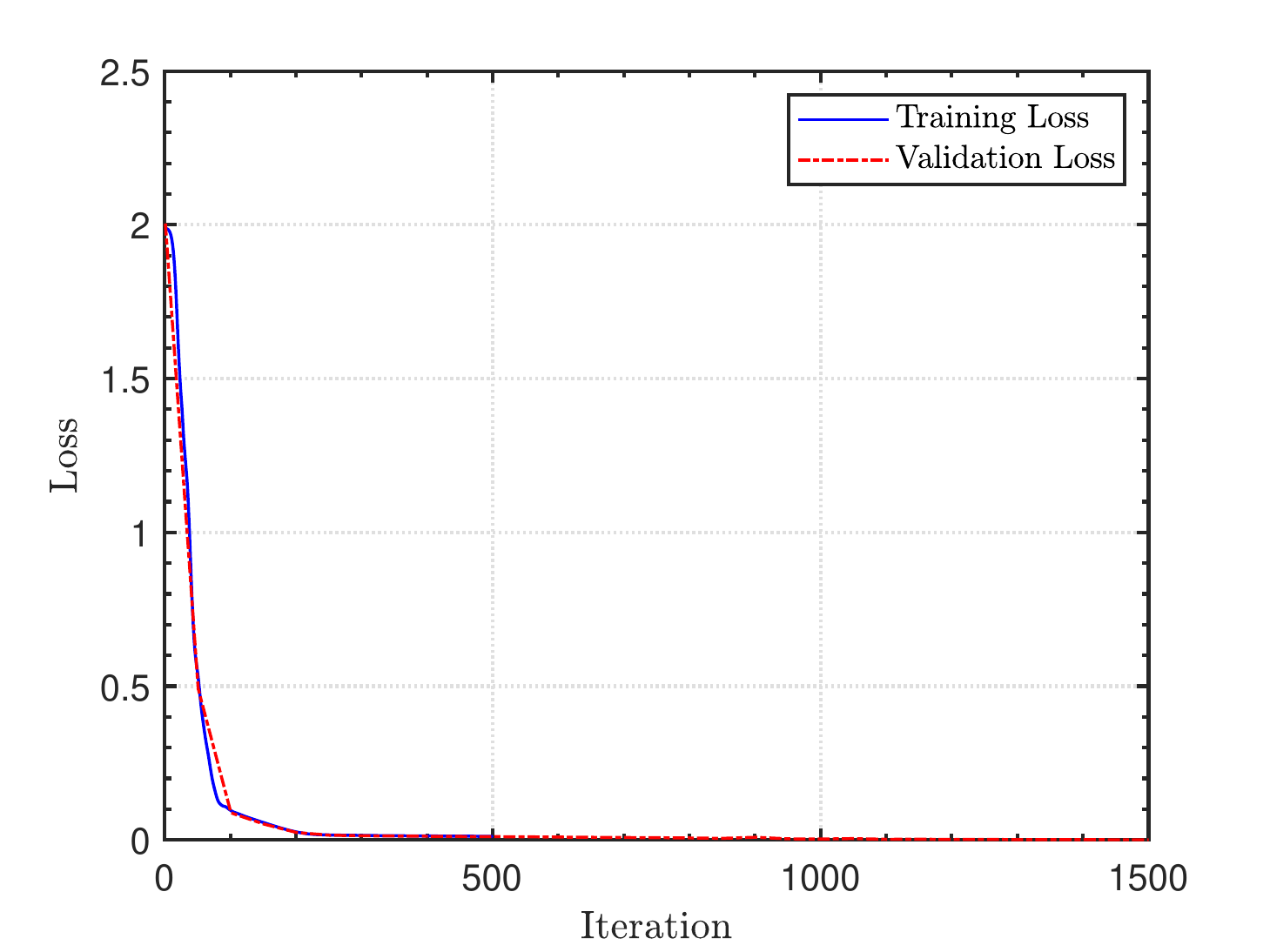}
		\caption{Loss function of the trained network with the parameters in Table \ref{Training_para}}
		\label{lossdnn}
		\vspace{-.2cm}
	\end{figure}

	We use a training set of size $10^5$ to learn the two predictor functions in \eqref{eq:dnn}.
The details about the training phase parameters are included in Table \ref{Training_para}. 
	\begin{table}[t!]
		\centering
		\caption{Training Parameters for the DNNs}
		\begin{tabular}{l*{6}{c}r}
			Parameter   & Value  \\
			\hline
			Number of batches & $10^4$ \\
			Size of batches      & 10  \\
			Number of epoches       &  2000 \\
			Number of iterations   &  $2\times10^7$
		\end{tabular}\label{Training_para}
	\vspace{-.4cm}
	\end{table}
Adam optimizer \cite{adam} with learning rate of $10^{-3}$ was used for loss function minimization. 
Fig. \eqref{lossdnn} compares the training loss and validation loss during the training phase at $20$ dB \ac{snr}. As seen, the gap between the training and validation loss diminishes when
the \ac{dnn} is trained
for more iterations.

	For a range of different \ac{snr}s and Doppler rates, we run $10^5$ Monte Carlo iterations to reach to a fair comparison between the existing channel estimator algorithms in terms of \ac{ber}.
	At each simulation setup we assume that the exact Doppler rate is known when DD-AR1 and DD-CC algorithms are employed while only the range of Doppler rate is known for the \ac{dl}-DD algorithm.
	
		\begin{figure}[]
		\vspace{-.2cm}
		\centering
		\includegraphics[width=0.5\textwidth]{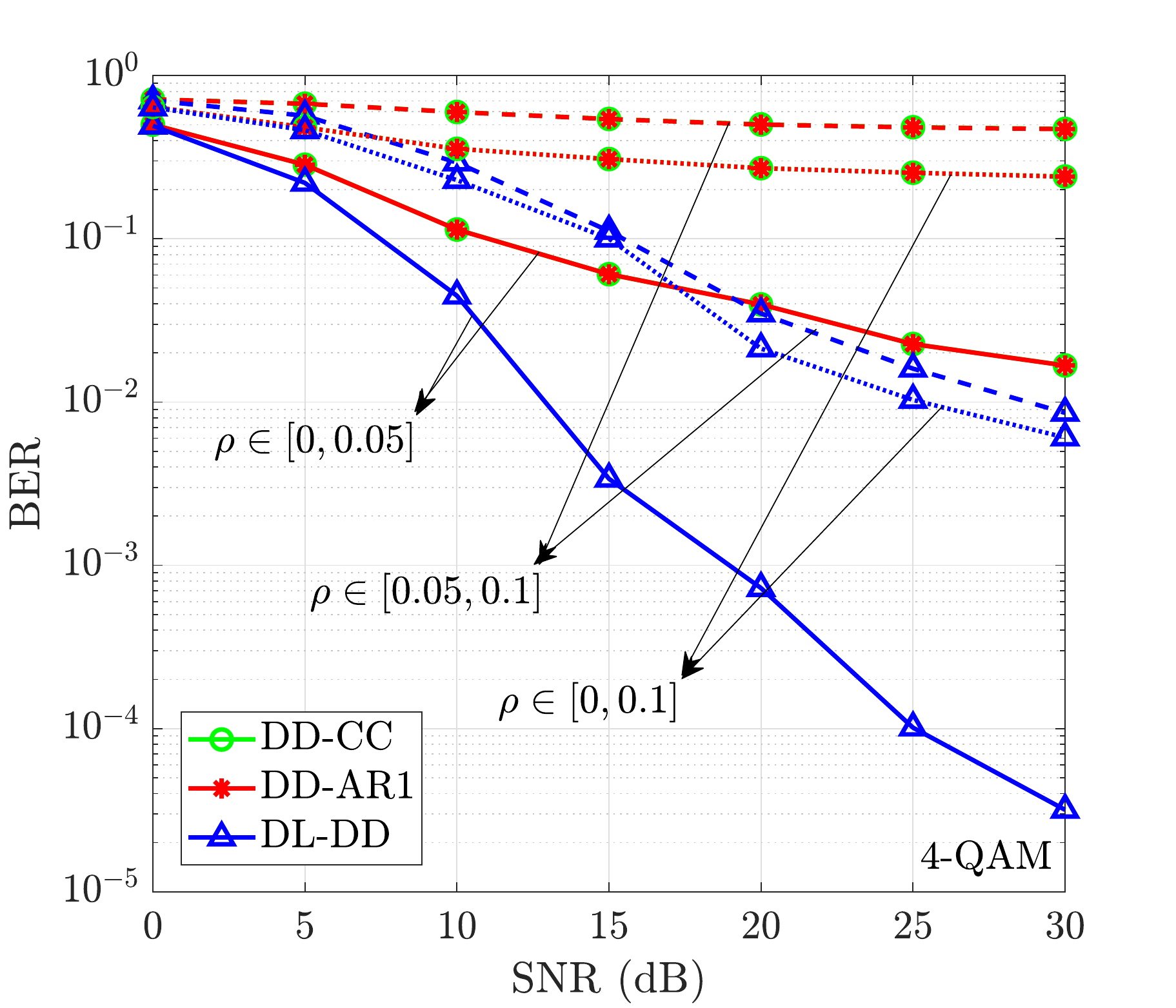}
		\caption{Comparison between the performance of \ac{dl}-DD, DD-AR1 and \ac{dd}-CC algorithms in terms of BER for different SNRs and range of Doppler rates, where Alamouti's \ac{stbc} \eqref{eq:alamouti} in a Rayleigh channel is used, $n_{\rm{p}}=10$, and $L=100$.}
		\label{fig_sim1}
		\vspace{-.6cm}
	\end{figure}
	
	\begin{figure}[]
		\vspace{-.5cm}
		\centering
		\includegraphics[width=0.5\textwidth]{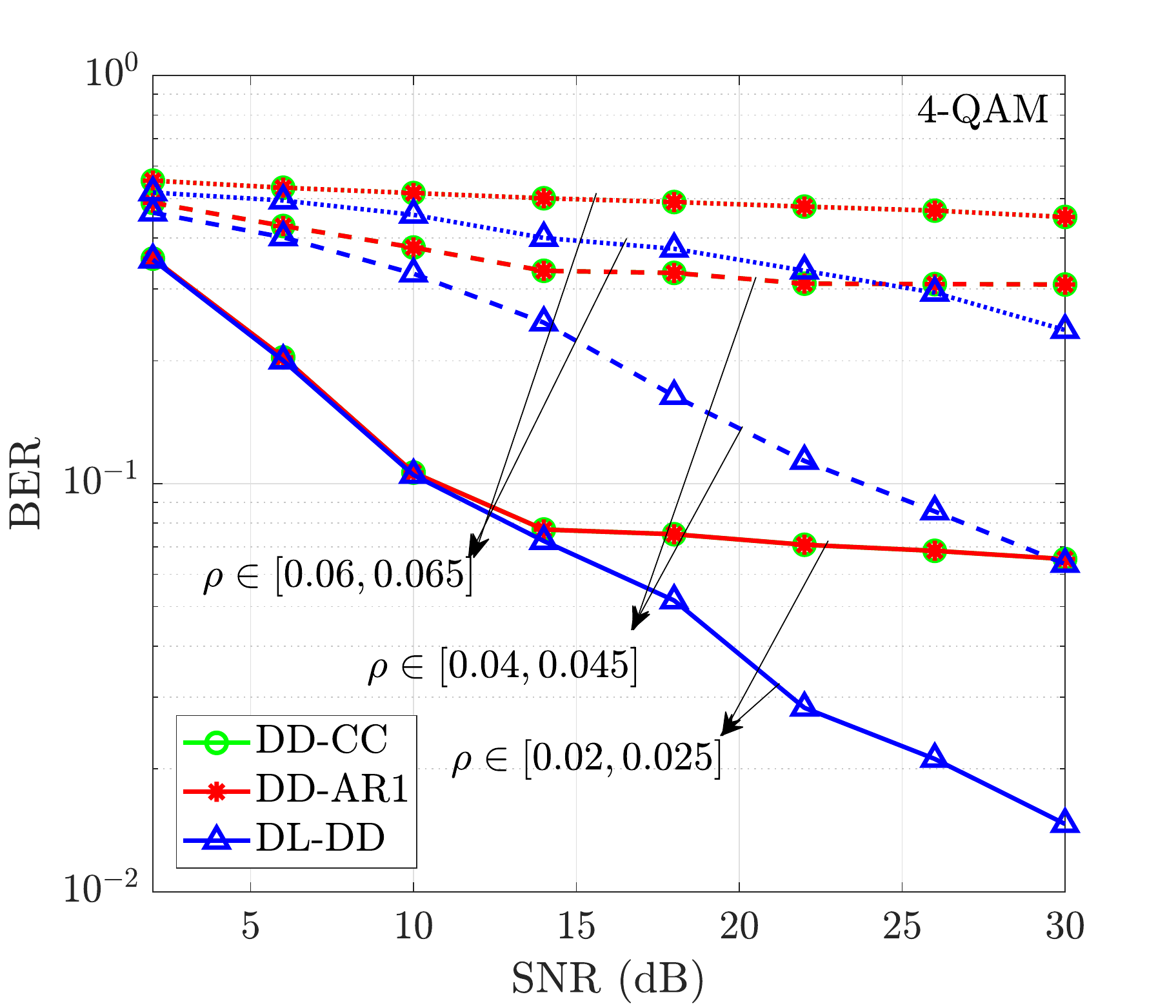}
		\caption{Comparison between the performance of \ac{dl}-DD, DD-AR1 and \ac{dd}-CC algorithms in terms of BER for different SNRs and range of Doppler rates, where Tarokh \textit{et. al.}'s \ac{stbc} \eqref{eq:tarokh} in a Rayleigh channel is used, $n_{\rm{p}}=10$, and $L=100$.}
		\label{fig_sim2}
		\vspace{-.6cm}
	\end{figure}
	
	\subsection{Simulation Results}
\begin{figure}[]
		\vspace{-.5cm}
		\centering
		\includegraphics[width=0.5\textwidth]{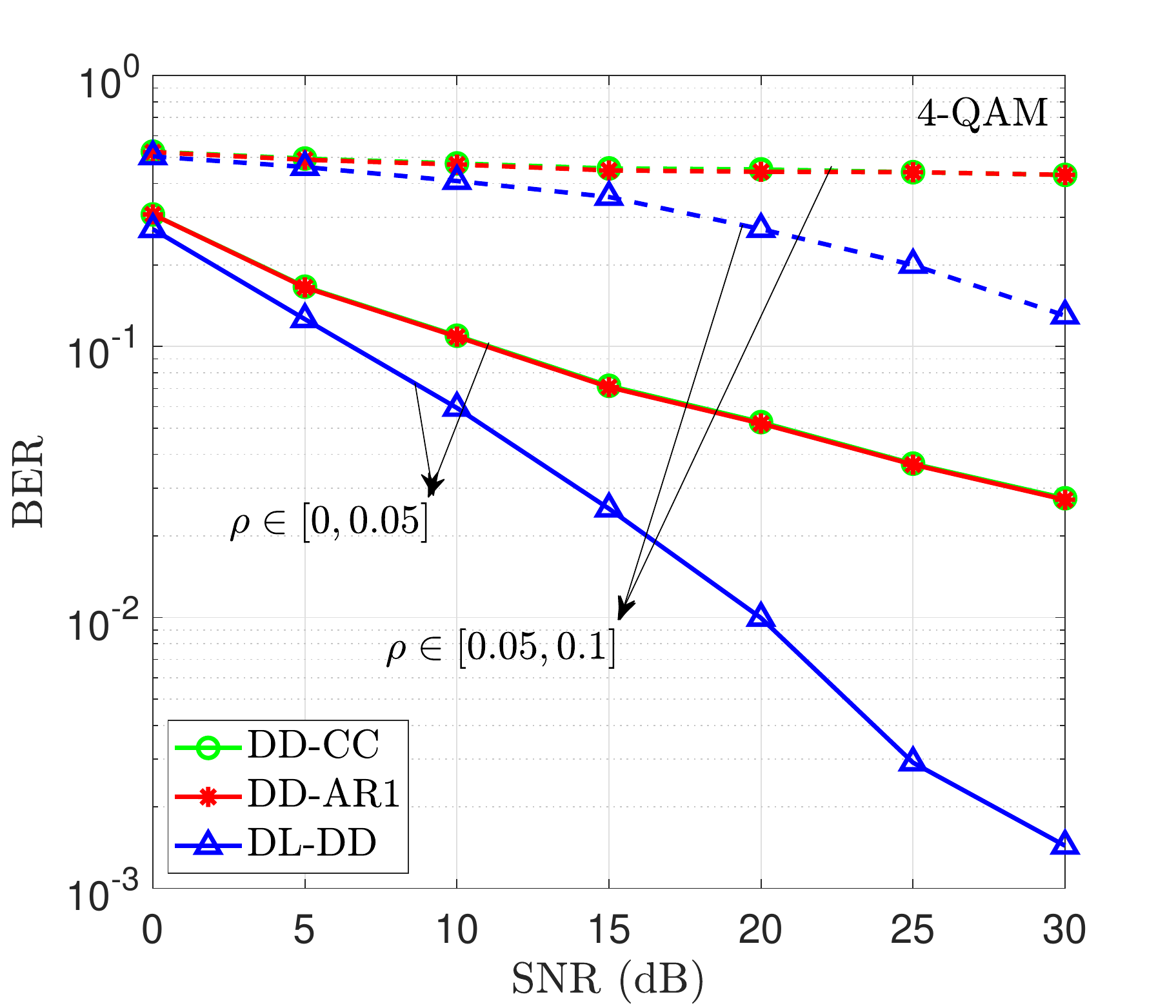}
		\caption{Comparison between the performance of \ac{dl}-DD, DD-AR1 and \ac{dd}-CC algorithms in terms of BER for different SNRs and range of Doppler rates, where the \ac{stbc} in \eqref{eq:tarokh343435} in a Rayleigh channel is used, $n_{\rm{p}}=10$, and $L=100$.}
		\label{fig_sim3}
		\vspace{-.5cm}
	\end{figure}

\begin{figure}[]
		\centering
		\includegraphics[width=0.5\textwidth]{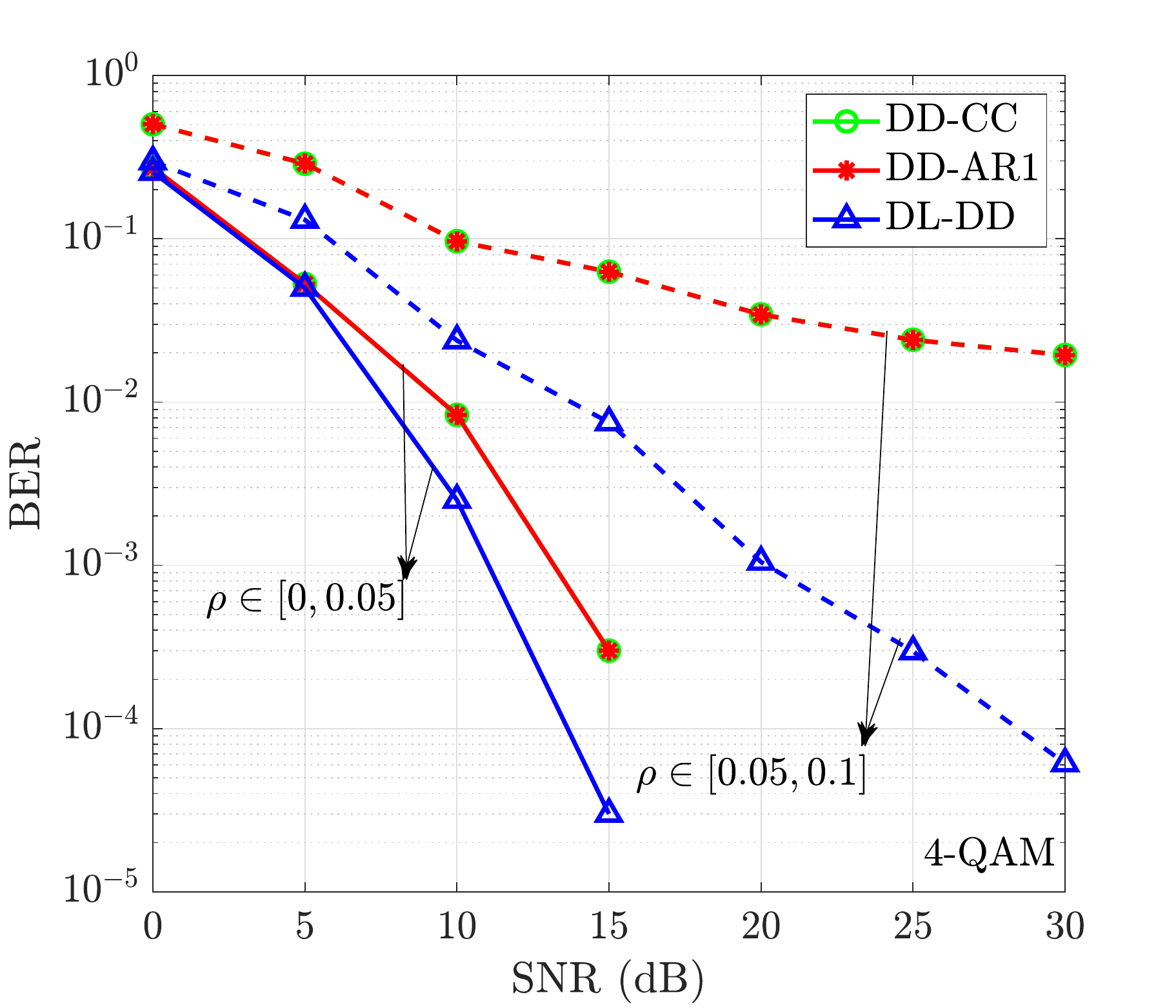}
		\caption{Comparison between the performance of \ac{dl}-DD, DD-AR1 and \ac{dd}-CC algorithms in terms of BER for different SNRs and range of Doppler rates, where Alamouti's \ac{stbc} \eqref{eq:alamouti} in a Racian channel with $k$-factor=2 is used, $n_{\rm{p}}=10$, and $L=100$.}
		\label{fig_sim4}
		\vspace{-.5cm}
	\end{figure}

	The performance of the \ac{dl}-DD, DD-AR1 and DD-CC algorithms have been studied, and they have employed in different ranges of Doppler rates.
	Fig.~\ref{fig_sim1}, Fig.~\ref{fig_sim2} and Fig.~\ref{fig_sim3} shows the performance comparison between these algorithms for Alamouti's \ac{stbc}, Tarokh \textit{et. al.}'s \ac{stbc} and \ac{stbc} in \eqref{eq:tarokh343435}, respectively for a Rayleigh channel.
	It is obvious from these figures that our proposed algorithm dramatically outperform the DD-AR1 and DD-CC algorithms at any \ac{snr}s and Doppler ranges in both cases even without the knowledge of the Doppler rate.
	As expected, increasing the \ac{snr} results in lower \ac{ber} and this reduction in \ac{ber} is more considerable in our algorithm.
	We repeat this simulation for a Rician channel with Alamouti's \ac{stbc} and provide the results in Fig. \ref{fig_sim4}.
	As seen, again our DL-DD algorithm outperforms the DD-AR1 and DD-CC algorithms.
	
\begin{figure}[]
		\vspace{-.5cm}
		\centering
		\includegraphics[width=0.5\textwidth]{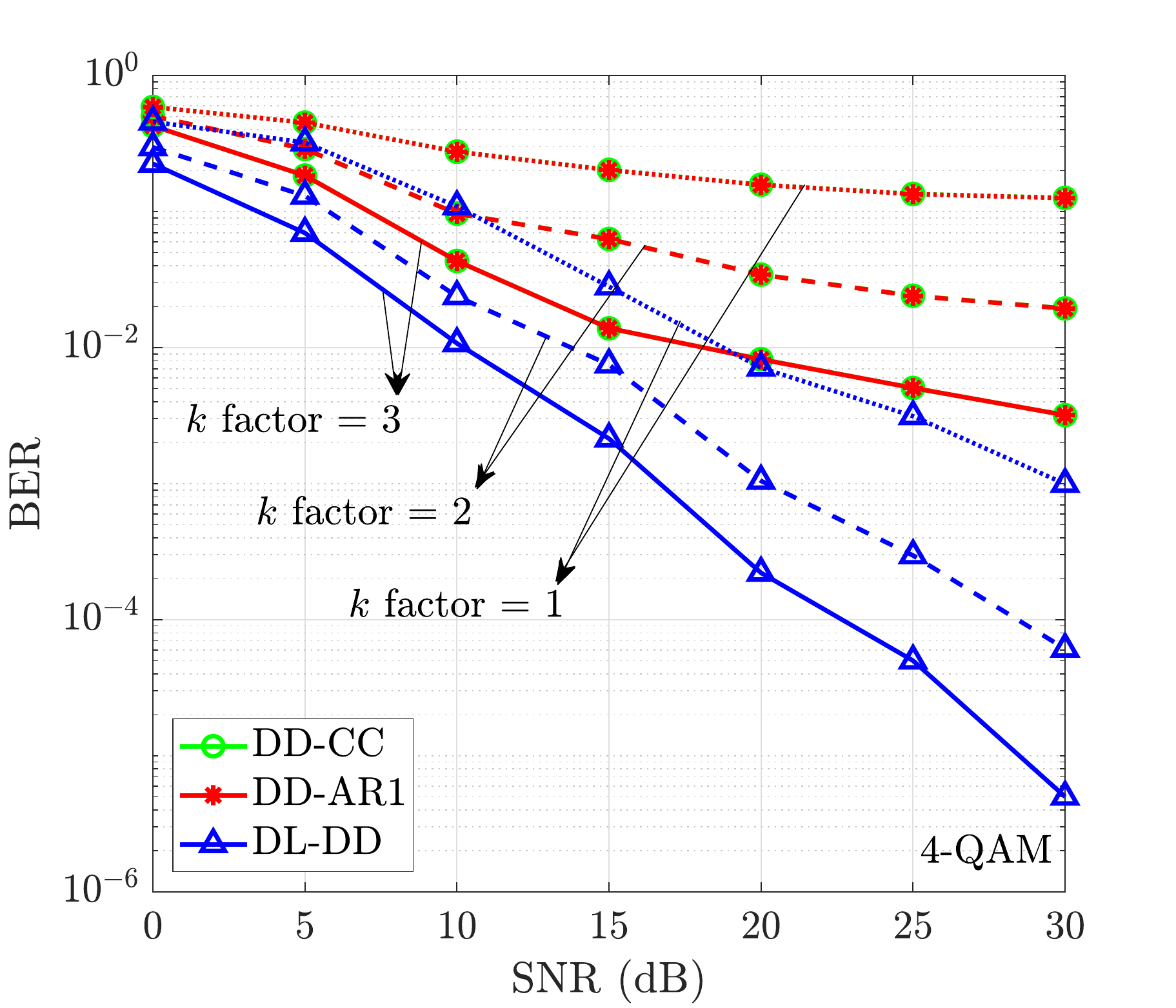}
		\caption{Comparison between the performance of \ac{dl}-DD, DD-AR1 and \ac{dd}-CC algorithms in terms of BER in three Rician channels with different $k$-factors for different SNRs, where Alamouti's \ac{stbc} \eqref{eq:alamouti} is used, $n_{\rm{p}}=10$, and $L=100$.}
		\label{fig_sim5}
	\end{figure}

	One of the parameters of a Rician channel that could affect the performance is $K$ \textit{factor}.
	We study the effect of $K$ factor on the achieved \ac{ber} by our DL-DD algorithm for Alamouti's \ac{stbc} and provide it in Fig. \ref{fig_sim5}.
	It is obvious from the figure that the performance of our DL-DD algorithm is considerably better than DD-CC and DD-AR1 algorithms and as the value of $k$-factor increases we obtain better \ac{ber}	with all the algorithms.

\begin{figure}[]
\vspace{-0.5cm}
		\centering
		\includegraphics[width=0.5\textwidth]{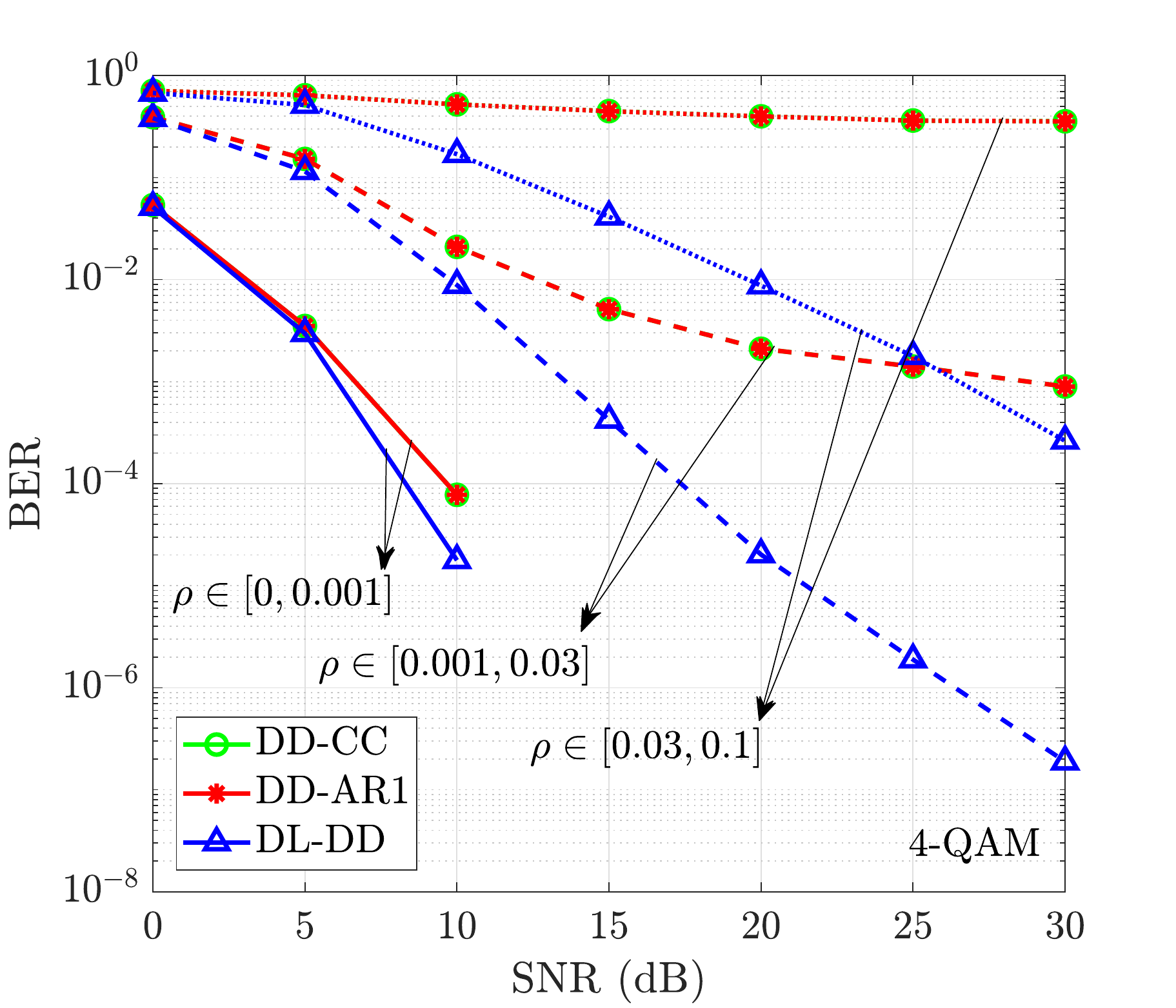}
		\caption{Comparison between the performance of \ac{dl}-DD, DD-AR1 and \ac{dd}-CC algorithms in terms of BER for different SNRs and three types of moving objects, where Alamouti's \ac{stbc} \eqref{eq:alamouti} in a Rayleigh channel is used, $n_{\rm{p}}=10$, and $L=100$.}
		\label{fig_sim6}
		\vspace{-.5cm}
	\end{figure}

In order to study the effect of moving object's speed on the performance of the channel predictors, we define three distinct Doppler rate ranges based on the speed of moving objects and provide a comparison in the following.
	We have the following equation for the relation between Doppler rate $\rho$ and moving object's speed $v$ as
	\begin{equation}
	\rho = \frac{vf_{\rm{c}}T_{\rm{c}}}{C},
	\end{equation}
	where $f_{\rm{c}}$ is the carrier frequency which is typically in the order of 10 GHz in 5G \cite{mmWave_5G_2}, $T_{\rm{c}}$ is the sampling time, and $C$ is the speed of light, i.e. $3\times10^8$~m/s.
	We consider three Doppler rate ranges for pedestrians, cars and high speed trains as in Table \ref{DopplerRanges}.
	\begin{table}[t!]
		\vspace{0.75cm}
		\centering
		\caption{List of Doppler rate ranges for different type of moving objects}
		\begin{tabular}{l*{6}{c}r}
			Name   & Speed (m/s) & Doppler Rate Range  \\
			\hline
			Pedestrians & $v\in[0,1]$ m/s & $\rho\in[0,0.001]$   \\
			Cars      & $v\in[1,60]$ m/s &  $\rho\in[0.001,0,03]$  \\
			High Speed Trains       &  $v\in[60,200]$ m/s &  $\rho\in[0,03,0.1]$
		\end{tabular}\label{DopplerRanges}
	\end{table}
	Fig.\ref{fig_sim6} shows the performance comparison between DL-DD, DD-AR1 and DD-CC for Alamouti's \ac{stbc} and the Doppler rate ranges in Table \ref{DopplerRanges}.
	As seen, our proposed DL-DD algorithm outperforms DD-AR1 and DD-CC in terms of \ac{ber}.
	
\begin{figure}[]
	\vspace{-.5cm}
		\centering
		\includegraphics[width=0.5\textwidth]{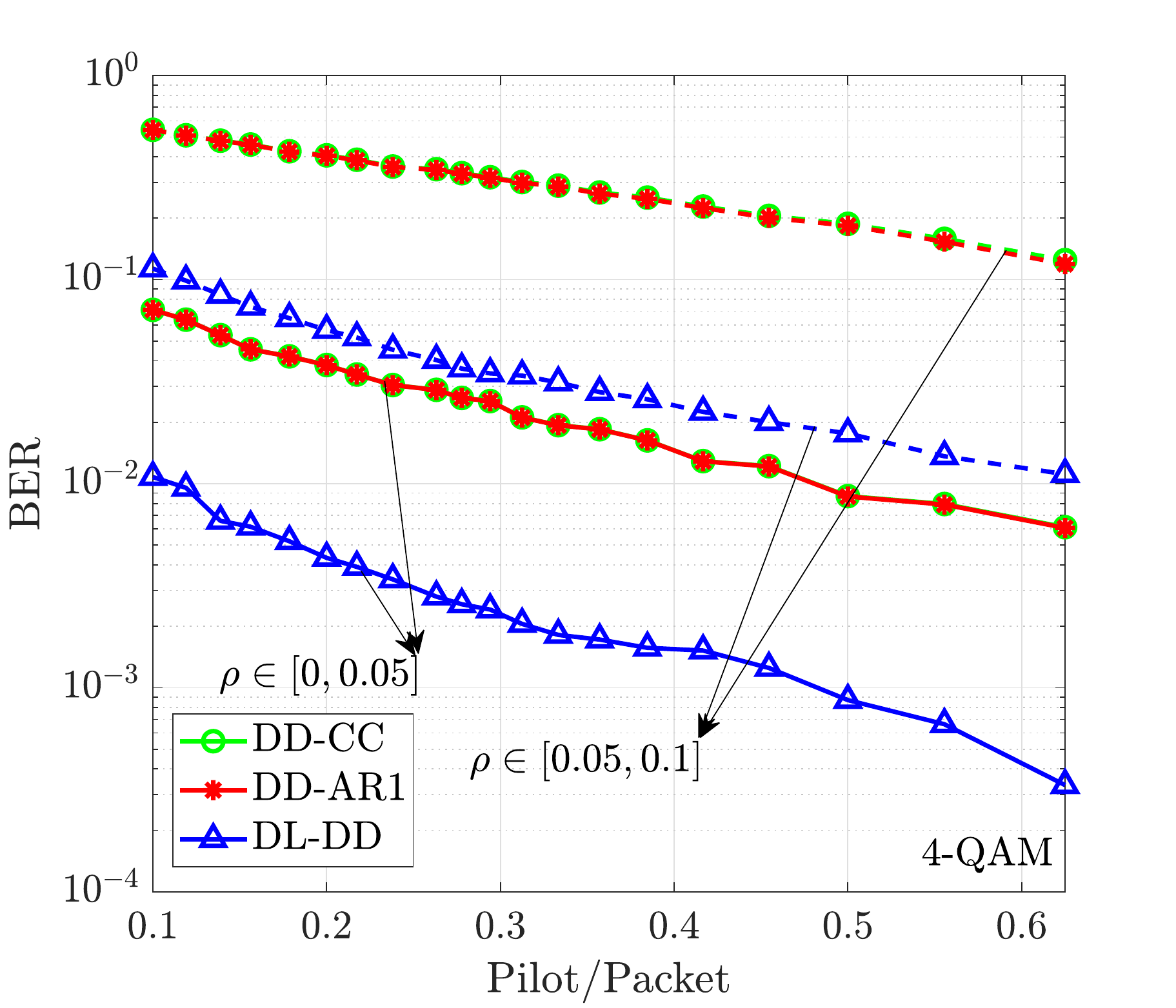}
		\caption{The effect of the packet length on the BER of the proposed \ac{dl}-DD, \ac{dd}-AR1 and DD-CC algorithms for different Doppler rate ranges at $15$~dB \ac{snr} in a Rayleigh channel.}
		\label{fig_rate}
	\end{figure}

	We study the effect of packet length on \ac{ber} and show the \ac{ber} versus $r=n_{\rm{p}}/L$ for $\rho \in [0 \ 0.05]$ and $\rho \in [0.05 \ 0.1]$ at $15$ dB for Alamouti's \ac{stbc} in Fig.~\ref{fig_rate}.
	It is assumed that the channel is in Rayleigh distribution and $n_{\rm{p}}=10$ and the number of \ac{stbc} transmission block, $n_{\rm{b}}$, varies.
	As seen, the proposed \ac{dl}-DD algorithm improves transmission reliability for long packets compared to the DD-AR1 and DD-CC algorithms.
	The reason is that the channel prediction error in the \ac{dl}-DD algorithm is much lower that the one in the other algorithms.
	The lower prediction error in the \ac{dl}-DD algorithm leads to lower propagation error.
	
\begin{figure}[]
	\vspace{-.5cm}
		\centering
		\includegraphics[width=0.5\textwidth]{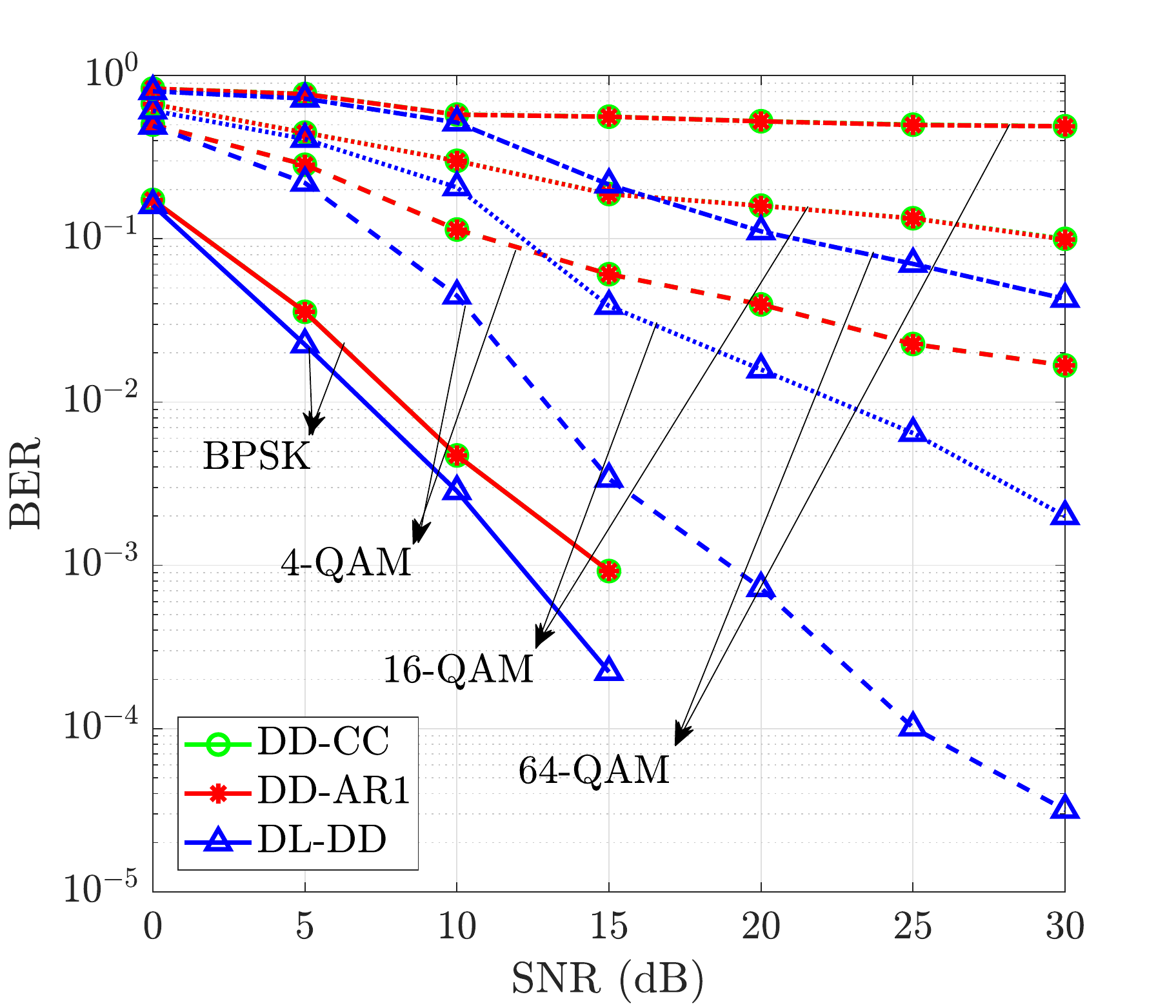}
		\caption{Comparison between the performance of \ac{dl}-DD, DD-AR1 and \ac{dd}-CC algorithms in terms of BER for different SNRs and three types of moving objects, where Alamouti's \ac{stbc} \eqref{eq:alamouti} in a Rayleigh channel is used, $n_{\rm{p}}=10$, and $L=100$.}
		\label{fig_sim7}
	\end{figure}

The effect of modulation format on the performance of the proposed \ac{dl}-based \ac{dd}-\ac{ce} algorithm for Alamouti's \ac{stbc} in Rayleigh fading channel is shown in Fig. \ref{fig_sim7}.
As seen, our proposed algorithm outperforms the other algorithms in terms of \ac{ber}. Also, as modulation order increases, the \ac{ber} increases.

\begin{figure*}[!h]
		\vspace{-.7cm}
		\begin{subfigure}[b]{0.5\textwidth}
			\includegraphics[width=9cm]{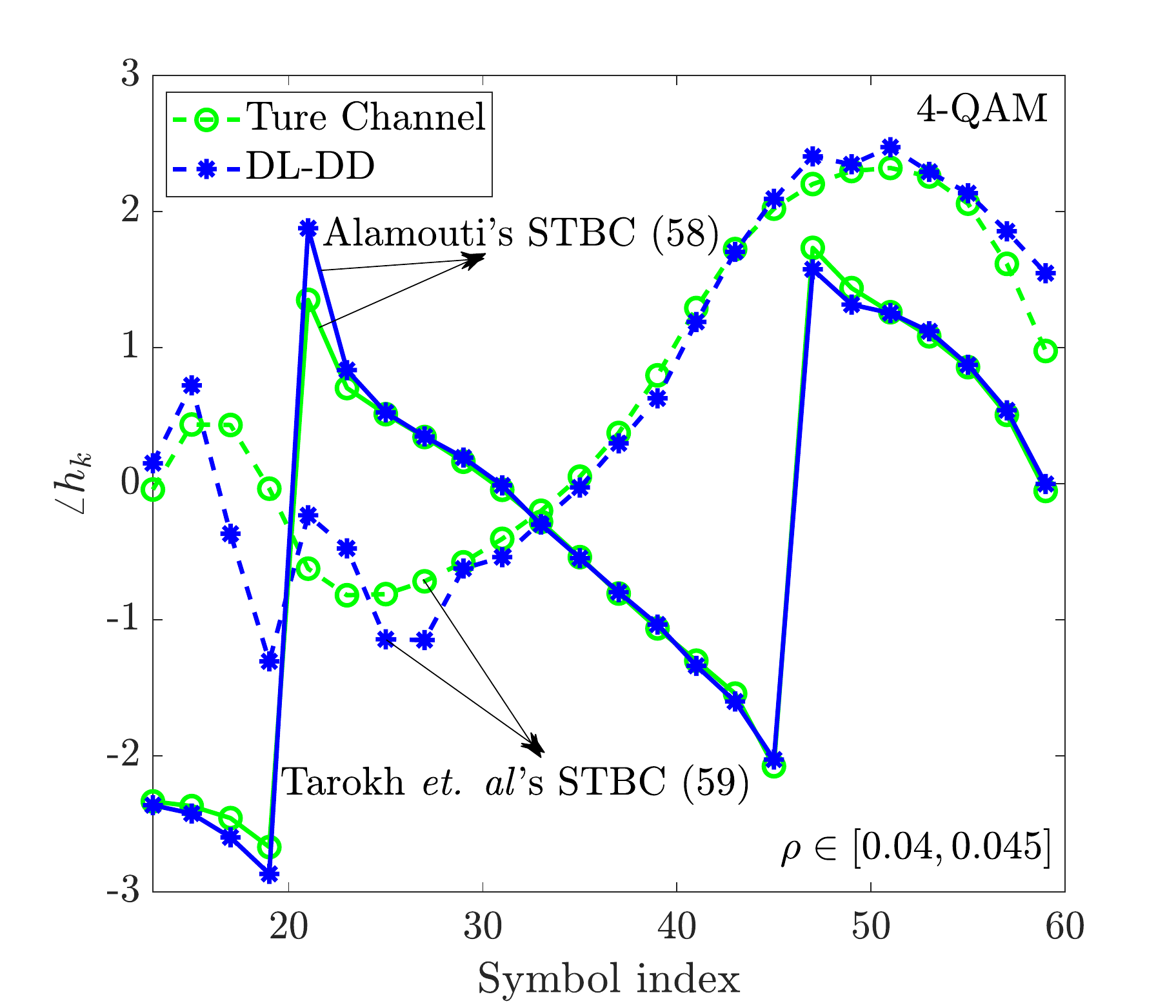}%
		\end{subfigure}
		\begin{subfigure}[b]{0.5\textwidth}
			\centering
			\includegraphics[width=9cm]{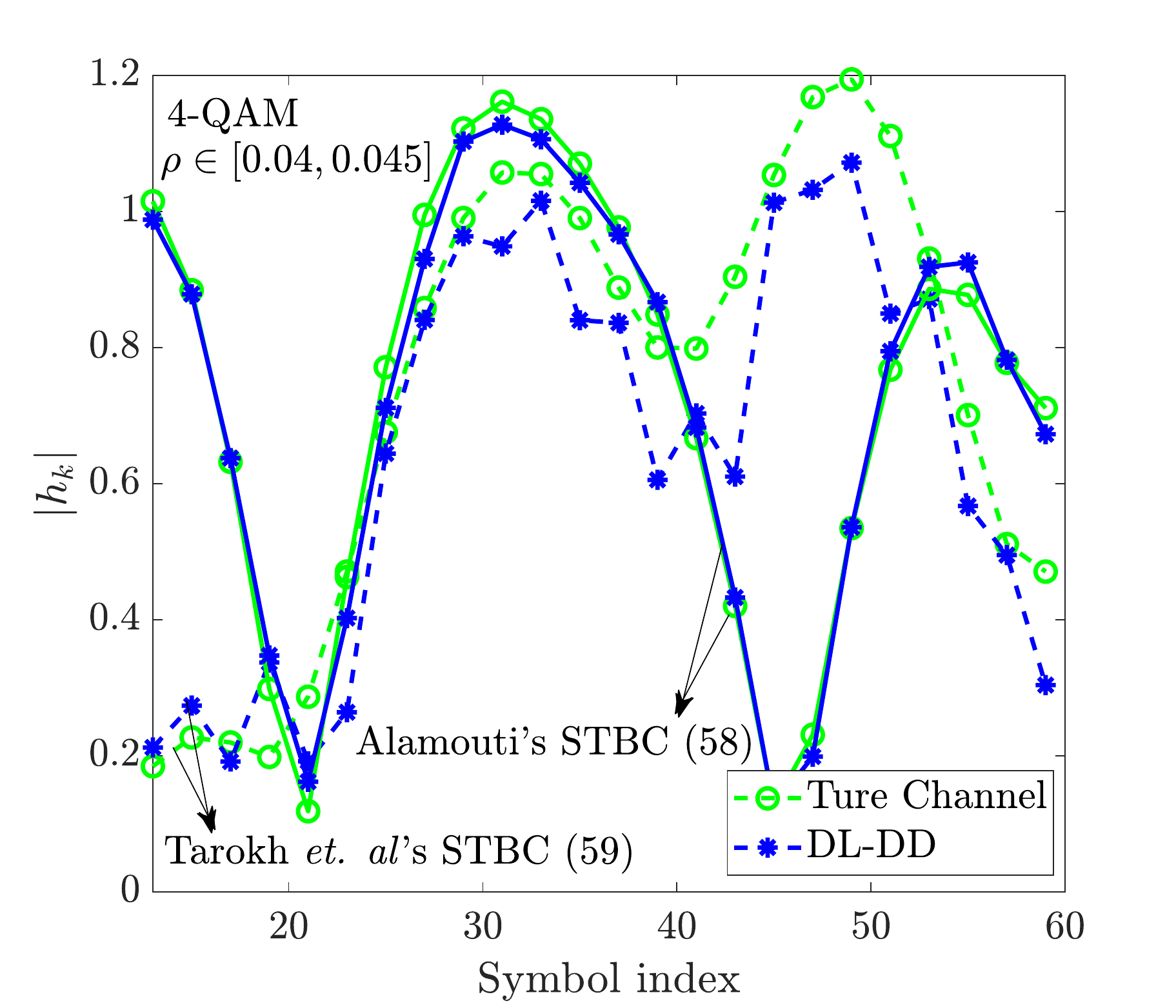}%
		\end{subfigure}
		\caption{Amplitude and phase tracking of the proposed DD-CC for Alamouti's \ac{stbc} \eqref{eq:alamouti} in a Rayleigh fading channel.}\label{fig_chtrac}
		\vspace{-.5cm}
	\end{figure*}

	Channel tracking capability of our proposed \ac{dl}-based algorithm in Rayleigh fading channel for  Alamouti's \ac{stbc} is presented in Fig.~\ref{fig_chtrac}.   
	As seen, the amplitude and phase of the predicted channels by the proposed \ac{dl}-DD algorithm is very close to the true channel for a packet transmission of length $L=100$.

	\section{Conclusion}\label{sec:con}
	The ac{mimo} communication systems enable us to achieve a higher data rate even in highly dynamic environments.
	However, this requires an improved \ac{ce} algorithm to be functional even in fast fading channels.
	In this paper we study \ac{dd}-\ac{ce} algorithm and develop a new \ac{dl}-based \ac{dd}-\ac{ce} algorithm to track fading channels and detect data for longer packets even in rapid vehicular environments.
	We also derive the \ac{ml} decoding formula for \ac{stbc} transmission.
	Our algorithm benefits from a simple receiver design which does not rely on the accurate statistical model of the fading channel and only the range of Doppler rate is sufficient.
	This capability removes the need for Doppler spread estimation, which is considerably challenging for highly dynamic vehicular environments.
	We compare our algorithm with \ac{dd}-AR1 and DD-CC algorithms through several performance measures and it outperforms existing algorithms while the DD-AR1 and DD-CC know the exact value of Doppler rate.
	
	\section*{Acknowledgment}
	The study presented in this paper is supported in part by the Huawei Innovation Research Program (HIRP).
	
\newpage
\bibliographystyle{IEEEtran}

\bibliography{IEEEabrv,Ref}

\end{document}